\documentclass[12pt]{iopart}
\usepackage{graphicx}
\begin{document}
\title {Quantum thermodynamics of a charged magneto-oscillator coupled to a heat bath.}
\vskip 0.5cm \author{Malay Bandyopadhyay}
\vskip 0.5cm
\address{Department of Theoretical Physics, Tata Institute of Fundamental Research, Homi Bhabha Road, Colaba, Mumbai-400005, India.}
\ead{malay@theory.tifr.res.in}
\begin{abstract}
\vskip 0.5cm
Explicit results for various quantum thermodynamic function (QTF) of a charged magneto-oscillator coupled to a heat bath at arbitrary temperature are demonstrated in this paper. Discernible expressions for different QTF in the two limits of very low and very high temperatures are presented for three popular heat bath models : Ohmic, single relaxation time and blackbody radiation. The central result is that the effect of magnetic field turns out to be important at low temperatures yet crucial at high temperatures. It is observed that the dissipation parameter, $\gamma$, and the cyclotron frequency, $\omega_c$, affect the decaying or rising behaviour of various QTF in just the opposite way to each other at low temperatures. In the high temperature regime, the effect of $\gamma$ is much pronounced than that of $\omega_c$. 
\end{abstract}
\maketitle
\section{Introduction}
\label{intro}
The problem of a charged quantum particle in the presence of a constant magnetic field is of great interest in the field of quantum Hall effect \cite{altshuler}, high temperature superconductivity \cite{hong}, diamagnetism \cite{van,malay1}, plasma physics \cite{a}, atomic physics \cite{b}, two dimensional electronic systems \cite{jacak}. The issue that we addres in this paper is what happens to the quantum thermodynamic functions (QTF) of a charged quantum particle in the presence of an external constant magnetic field when it is in contact with a dissipative quantum heat bath. This kind of analysis is related with the dissipative quantum mechanics, a subject that has seen a great attention through the work of Leggett and others \cite{legg1,legg2,legg3}. There are several approaches for the treatment of dissipative quantum systems. The most convensional approach is system plus reservoir point of view i.e. the system of interest is coupled linearly with the environment which is represented by a collection of harmonic oscillators \cite{li1,li2}. Usually, one is interested in the dissipative subsystem and the reservoir variables are eliminated by projection operator or tracing procedure \cite{c}. As a result of that, the reservoir enters only through few parameters. The results obtained from these kind of dissipative quantum systems are of great interest due to the recent widespread interest on the critical role of environmental effects in mesoscopic systems \cite{mazenko,datta,imry,chakravarty,tvr}, in fundamental quantum physics, and in quantum information \cite{bennett,zeilinger,giulini,myatt}. In last few years, research on the open quantum systems has questioned about validity of fundamental laws of thermodynamics \cite{capek,shicka,sheehan,allah}. These subtle issues are discussed in details by several authors \cite{hanggi1,hanggi2,hanggi3,ford2,kim1,kim2,bao,malayb}. Here, our goal is not to survey all these subtle issues, but we focus on the effect of environment and the effect of magnetic field on various thermodynamic funtions such as free enrgy, entropy, internal energy, and specific heat of a charged oscillator. By this realistic calculations, one can make direct contact with experiment on low dimensional nanostructures and  quantum dots \cite{jacak}.\\
\indent
Starting point of this paper is the famous Caldeira-Leggatt heat-bath model to incorporate the environmental effect in quantum and mesoscopic systems \cite{legg1,legg2}. This enables us to derive generalized quantum Langevin equation (GQLE) for the charged magneto-oscillator. Then, we use the previously derived result of Ford {\it et al} \cite{ford4} for the free energy of a charged magneto-oscillator in an arbitrary heat bath in terms of a single integral involving generalized susceptibility coming up from the generalized quantum Langevin equation \cite{ford1,o'connell}. This enables us to derive the free energy of the model system at an arbitrary temperature. Using this generalized result, one can obtain free energy expression and other thermodynamic functions of the charged quantum magneto-oscillator for three popular heat bath models : Ohmic, single relaxation time and blackbody radiation heat bath or Quantum Electrodynamics (QED) model.\\
\indent
Now, we need to verify in which way our results are new from the previous results in literature. Ford and O'Connell discussed about the quantum thermodynamic functions for an oscillator coupled to a heat bath \cite{ford5}. In this paper, we extend the work of Ford and O'Connell to include the presence of an external static magnetic field. We determine various quantum thermodynamic functions of the charged magneto-oscillator coupled to a heat bath. What we find that each term in the low temperature expansion of different QTF are added up by two additional magnetic field dependent terms. On the other hand, the inclusion of magnetic field in the problem is reflected in the two additional magnetic field dependent terms in each quantum thermodynamic function in the high temperature regime. In the low temperature regime, the magnetic field changes the qualitative behaviour as well as the quantitative values of different QTF for the charged magneto-oscillator  from that of a free oscillator. In this low temperature regime, $\omega_c$ and $\gamma$ affect deacaying or rising behaviour of various QTF in the opposite manner. It is seen that the effect of $\omega_c$ is negligible in the high temperature regime. But, the effect of $\gamma$ is still crucial at high temperatures. We also plot the general expression of various QTF for the entire temperature regime by numerically evaluating equation (24).\\
\indent
With this preceding background, we organize the rest of the paper as follows. In section II, we describe our model and the method of deriving free energy of the system. Section III is devoted to derive various thermodynamic functions (such as entropy, internal energy, specific heat) for the low temperature as well as high temperature regime for the Ohmic heat bath. On the other hand, we derive QTF for the single relaxation time and blackbody radiation heat bath in section IV. Zero temperature results are shown in section V. We conclude in section VI.\\
\section{Model \& Free energy}
The starting point of this section is the generalized Caldeira-Legget system-plus-reservoir Hamiltonian for a charged particle of mass `$m$' and charge `$e$' in a magnetic field $\vec{B}$ in the operator form \cite{legg1,legg2} :
\begin{eqnarray}
\hat{H}&=&\frac{\Big(\hat{p}-e\vec{A}/c\Big)^2}{2m}+V(\hat{r})+\sum_{j=1}^N\Big\lbrack\frac{\hat{p}_j^2}{2m_j}+\frac{1}{2}m_j\omega_j^2\Big(q_j-\frac{c_j}{m_j\omega_j}\hat{r}\Big)^2\Big\rbrack,
\end{eqnarray}
where $\lbrace\hat{r},\hat{p}\rbrace$ and $\lbrace\hat{q}_j,\hat{p}_j\rbrace$ are the sets of co-ordinate and momentum operators of system and bath oscillators. They follow the following commutation relations
\begin{eqnarray}
\lbrack \hat{r}_{\alpha},\hat{p}_{\beta}\rbrack = i\hbar\delta_{\alpha\beta},
\lbrack\hat{q}_{i\alpha},\hat{p}_{j\beta}\rbrack=i\hbar\delta_{ij}\delta_{\alpha\beta},
\end{eqnarray}
where Greek indices $\alpha$, $\beta$ stand for three spatial directions. Eliminating the bath degrees of freedom by the Heisenberg equations of motion one can obtain the generalized quantum Langevin equation (GQLE) \cite{malayb,ford1}:
\begin{eqnarray}
m\ddot{\hat{r}}+\int_{-\infty}^tdt^{\prime}\gamma(t-t^{\prime})\dot{\hat{r}}(t^{\prime})-\frac{e}{c}\dot{\hat{r}}\times\vec{B}+\vec{\nabla}V(\hat{r})=\hat{\theta}(t),
\end{eqnarray}
where dot denotes differentiation with respect to time $t$. The effect of the magnetic field is solely represented by the quantum version of the Lorentz force (third term in Eq. (3)). The memory kernel $\gamma(t)$ and the operator valued random force $\hat{\theta}(t)$ are unaffected by the magnetic field. In this work, we consider the confining potential to be harmonic for which an exact analysis is possible. It is now possible to represent nonequal time anticommutator and commutator of $\hat{\theta}(t)$ as follows:
\begin{eqnarray}
\langle\lbrace\hat{\theta}_{\alpha}(t),\hat{\theta}_{\beta}(t^{\prime})\rbrace\rangle&=&\delta_{\alpha\beta}\frac{\beta\hbar}{\pi}\int_0^{\infty}d\omega J(\omega)\coth\Big(\frac{\beta\hbar\omega}{2}\Big)\cos\lbrack\omega(t-t^{\prime})\rbrack,
\end{eqnarray}
\begin{equation}
\langle\lbrack\hat{\theta}_{\alpha}(t),\hat{\theta}_{\beta}(t^{\prime})\rbrack\rangle=\delta_{\alpha\beta}\frac{\beta\hbar}{\pi}\int_0^{\infty}d\omega J(\omega)\sin\omega(t-t^{\prime}),
\end{equation}
where $\beta=\frac{1}{k_BT}$ is the inverse temperature. The memory kernel is given by
\begin{equation}
\gamma(t)=\frac{2}{m\pi}\int_0^{\infty}d\omega \frac{J(\omega)}{\omega}\cos(\omega t).
\end{equation} 
In equations (4)-(6), $J(\omega)$ denotes the spectral density function of the heat bath oscillators and is given by :
\begin{eqnarray}
J(\omega)=\pi\sum_{j=1}^N\frac{c_j^2}{2m_j\omega_j}\delta(\omega-\omega_j)
\end{eqnarray}
We are interested in investigating thermodynamic behaviour of a dissipative charged magneto-oscillator at an arbitrary temperature. One can easily determine the free energy for this model system by using the remarkable formula \cite{ford4}
\begin{eqnarray}
F =\frac{1}{\pi}\int_0^{\infty}d\omega f(\omega,T)\Im\Big\lbrack\frac{d}{d\omega}\ln\Big(\det\alpha(\omega+i0^+)\Big)\Big\rbrack,
\end{eqnarray}
where $f(\omega,T)$ is the free energy of a single oscillator of frequency $\omega$ and is given by 
\begin{equation}
f(\omega,T)=k_BT\log\Big\lbrack 1 - \exp(-\frac{\hbar\omega}{k_BT})\Big\rbrack,
\end{equation}
where we have ignored the zero-point contribution which is discussed in section V. Here $\alpha(\omega)$ denotes the generalized susceptibility of the model system.  Since all the results presented in this paper rely on this remarkable formula, it is the cornerstone of the paper. So, we are giving a sketch of the way of deriving this notable formula. It is known that the free enrgy ascribed to the magneto-oscillator, F(T,B), is the free energy of the magneto-oscillator coupled to the heat bath minus the free energy of the bath in the absence of the magneto-oscillator. To derive this free energy one can follow the following method. Using Fourier transform one can rewrite equation (3) as follows :
\begin{equation}
\tilde{r}(\omega)=\alpha(\omega)\tilde{\theta}(\omega),
\end{equation}
where, $\tilde{r}(\omega)$ and $\tilde{\theta}(\omega)$ are the Fourier transform of the operators $\hat{r}(t)$ and $\hat{\theta}(t)$ respectively. Now, it can be easily shown that $\alpha(\omega)$ have poles on the real axis at the normal mode frequencies, $\tilde{\omega}_j$, of the interacting system and zeroes at the bath frequencies, $\omega_i$, in the absence of the charged magneto oscillator. Therefore, one can write
\begin{equation}
\alpha(\omega)=-\frac{1}{m}\frac{\prod_i (\omega^2-\omega_i^2)}{\prod_j(\omega^2-\tilde{\omega}_j^2}),
\end{equation}
where the numerator is the product over normal modes of the free bath oscillators and the denominator is the product over those of the interacting system. From the well known formula $\frac{1}{(x+i0^+)}=P(1/x)-i\pi\delta(x)$, one can show that :
\begin{equation}
\hskip-1.0 cm
\frac{1}{\pi}\Im\lbrack\frac{d}{d\omega}\ln\alpha(\omega)\rbrack=\sum_j\lbrack\delta(\omega-\tilde{\omega}_j)+\delta(\omega+\tilde{\omega}_j)\rbrack - \sum_i\lbrack\delta(\omega-{\omega}_i)+\delta(\omega+{\omega}_i)\rbrack.
\end{equation}
When this is put into equation (8), it results :
\begin{equation}
F(T,B)=\sum_j f(\tilde{\omega}_j, T)-\sum_i f(\omega_i,T),
\end{equation}
where, the first sum is clearly the free energy of the interacting system and second that of the free bath field. This demonstrates our assertion. Now, we can rewrite Eq. (8) as follows \cite{li1,li2}:
\begin{equation}
F(T,B)=F(T,0)+\Delta F(T,B),
\end{equation}
where 
\begin{equation}
F(T,0)=\frac{3}{\pi}\int_0^{\infty}d\omega f(\omega, T)I_1
\end{equation}
is the free energy of the oscillator in the absence of the magnetic field, $I_1=\Im\Big\lbrack\frac{d}{d\omega}\ln\alpha^{(0)}(\omega)\Big\rbrack$, $\alpha^{(0)}(\omega)$ is the scalar susceptibility in the absence of a magnetic field and the correction due to the magnetic field is given by
\begin{equation}
\Delta F(T,B)=-\frac{1}{\pi}\int_0^{\infty}d\omega f(\omega, T)I_2,
\end{equation}
where $I_2=\Im\Big\lbrace\frac{d}{d\omega}\ln\Big\lbrack 1-\Big(\frac{eB\omega\alpha^{(0)}}{c}\Big)^2\Big\rbrack\Big\rbrace$. The scalar susceptibility for a harmonic oscillator in the absence of a magnetic field is given by \cite{li1,li2}:
\begin{equation}
\alpha^{(0)}(\omega)=\frac{1}{m(\omega_0^2-\omega^2)-i\omega\tilde{\gamma}(\omega)},
\end{equation}
where
\begin{equation}
\tilde{\gamma}(\omega)=\int_0^tdt^{\prime}\gamma(t^{\prime})e^{i\omega t^{\prime}}.
\end{equation}
We have now all the essential ingredients to calculate thermodynamic functions. Our main task is to find free energy $F$. Then, one can easily derive other thermodynamic functions at an arbitrary temperature.
\begin{figure}[t]
\begin{center}
{\rotatebox{0}{\resizebox{12cm}{6cm}{\includegraphics{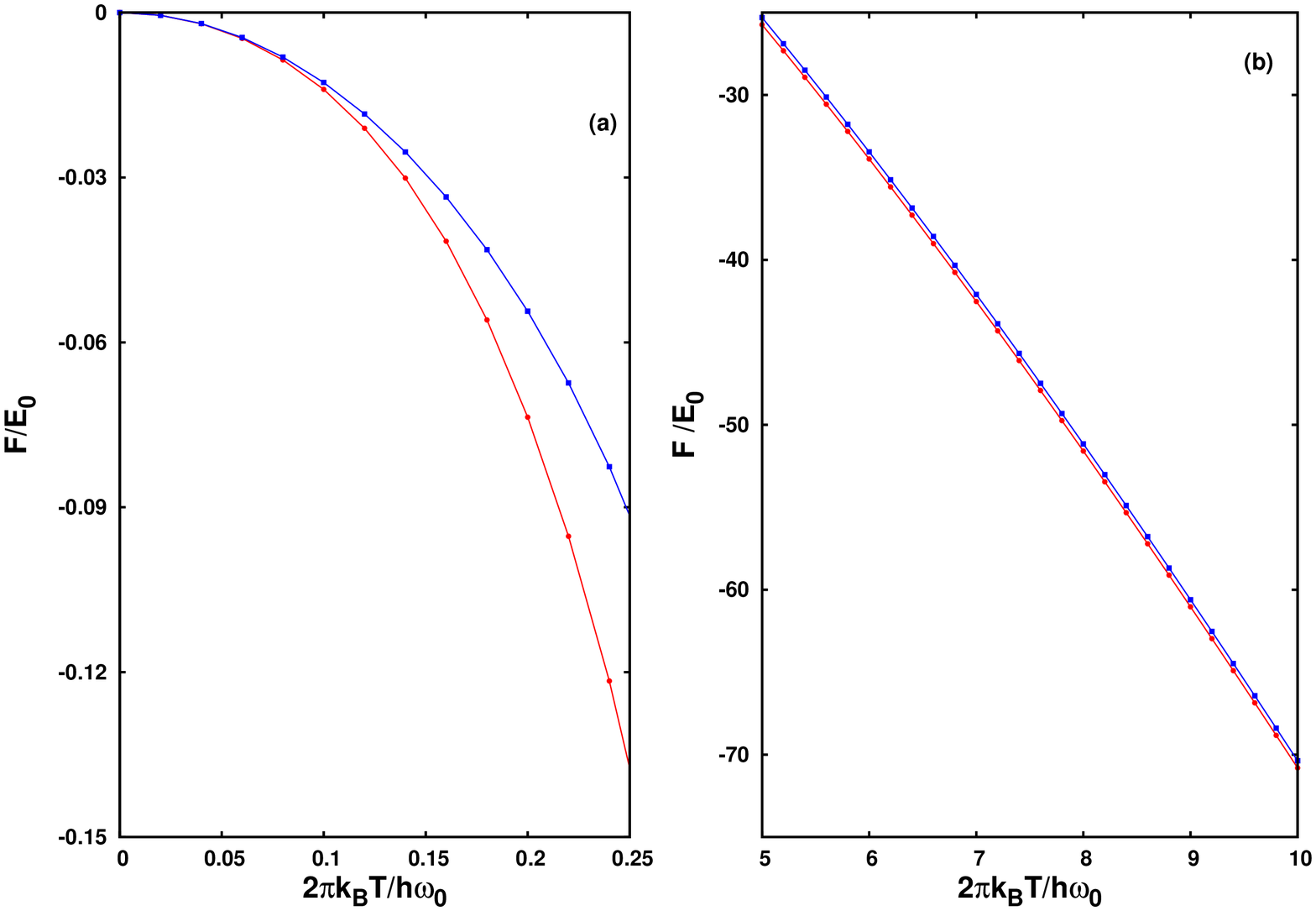}}}}
%\vskip-2.5cm
\caption{(color online)  Plot of $\frac{F}{E_{0}}$ versus dimensionless temperature, $\frac{2\pi k_BT}{h\omega_0}$, for the charged magneto-oscillator  (red filled circle) and for the free  charged oscillator (blue filled square) coupled to a Ohmic or SRT heat bath (a) in the low temperature regime and (b) in the high temperature regime. To plot this figure, we use $\frac{\gamma}{\omega_0}=0.8$, $\frac{\gamma}{\omega_c}=1.6$, and $\frac{\omega_0}{\omega_c}=2.0$.} 
\end{center}
\end{figure} 
 Entropy is defined as 
\begin{equation}
S(T,B)=-\frac{\partial F(T,B)}{\partial T}.
\end{equation}
Internal energy is given by
\begin{equation}
U(T,B)=F(T,B)+TS(T,B),
\end{equation}
and specific heat is defined as 
\begin{equation}
C(T,B)=T\frac{\partial S(T,B)}{\partial T}.
\end{equation}
Now, we can proceed for the three popular heat-bath models of our interest which are characterized by the following spectral density functions (a) Ohmic heat bath : $\tilde{\gamma}(\omega)=\gamma_0$, (b) single relaxation time : $\tilde{\gamma}(\omega)=\frac{\gamma_0}{1-i\omega\tau}$, and (c) blackbody radiation : $ \tilde{\gamma}(\omega)= \frac{2e^2\omega\Omega^2}{3c^3(\omega+i\Omega)}$, where $\gamma_0$ is the Ohmic friction constant, while $\tau$ is the relaxation time. Usually $\tau$ is small i.e. $\tau<<\frac{\gamma_0}{m}$. $\Omega$ is the high frequency cut-off characterizing electron form factor for blackbody radiation model. One can easily express $\alpha^{(0)}(\omega)$ for all the three cases by a single expression \cite{ford5} :
\begin{equation}
\alpha^{(0)}(\omega)=\frac{\omega+i\Omega}{m(\omega+i\Omega^{\prime})(\omega_0^2-\omega^2-i\gamma\omega)}.
\end{equation}
Following Ford and O'Connell, one can write $\tau=\frac{1}{\Omega}=\frac{1}{\Omega^{\prime}+\gamma}$; $\frac{\gamma_0}{m}=\gamma\frac{{\Omega^{\prime}}^2+\gamma\Omega^{\prime}+\omega_0^2}{(\Omega^{\prime}+\gamma)^2}$; $\frac{K}{m}=\omega_0^2\frac{\Omega^{\prime}}{\Omega^{\prime}+\gamma}$ for the single relaxation time model \cite{ford5}. The Ohmic model can be represented by $\tau\rightarrow 0$, $\frac{\gamma_0}{m}\rightarrow \gamma$ and $\frac{K}{m}\rightarrow \omega_0^2$. On the other hand, for the blackbody radiation heat bath or QED model, $\frac{1}{\Omega}=\frac{1}{\Omega^{\prime}}+\frac{\gamma}{\omega_0^2}$; $\frac{K}{M}=\omega_0^2\frac{\Omega^{\prime}}{\Omega^{\prime}+\gamma}$; $\frac{M}{m}=\frac{(\omega_0^2+\gamma\Omega^{\prime})(\Omega^{\prime}+\gamma)}{\omega_0^2\Omega^{\prime}}$, where $m$ is the bare mass and $M=m+\frac{2e^2\Omega}{3c^3}$ is the renormalized mass. In the largest value of cut-off limit $(\Omega^{\prime}\rightarrow \infty)$, $m=0$, $K=M\omega_0^2$ and $\Omega =\frac{1}{\tau_e}$, where $\tau_e=\frac{2e^2}{3Mc^3}=6\times10^{-24} s$.
\begin{figure}[t]
\begin{center}
{\rotatebox{0}{\resizebox{12cm}{6cm}{\includegraphics{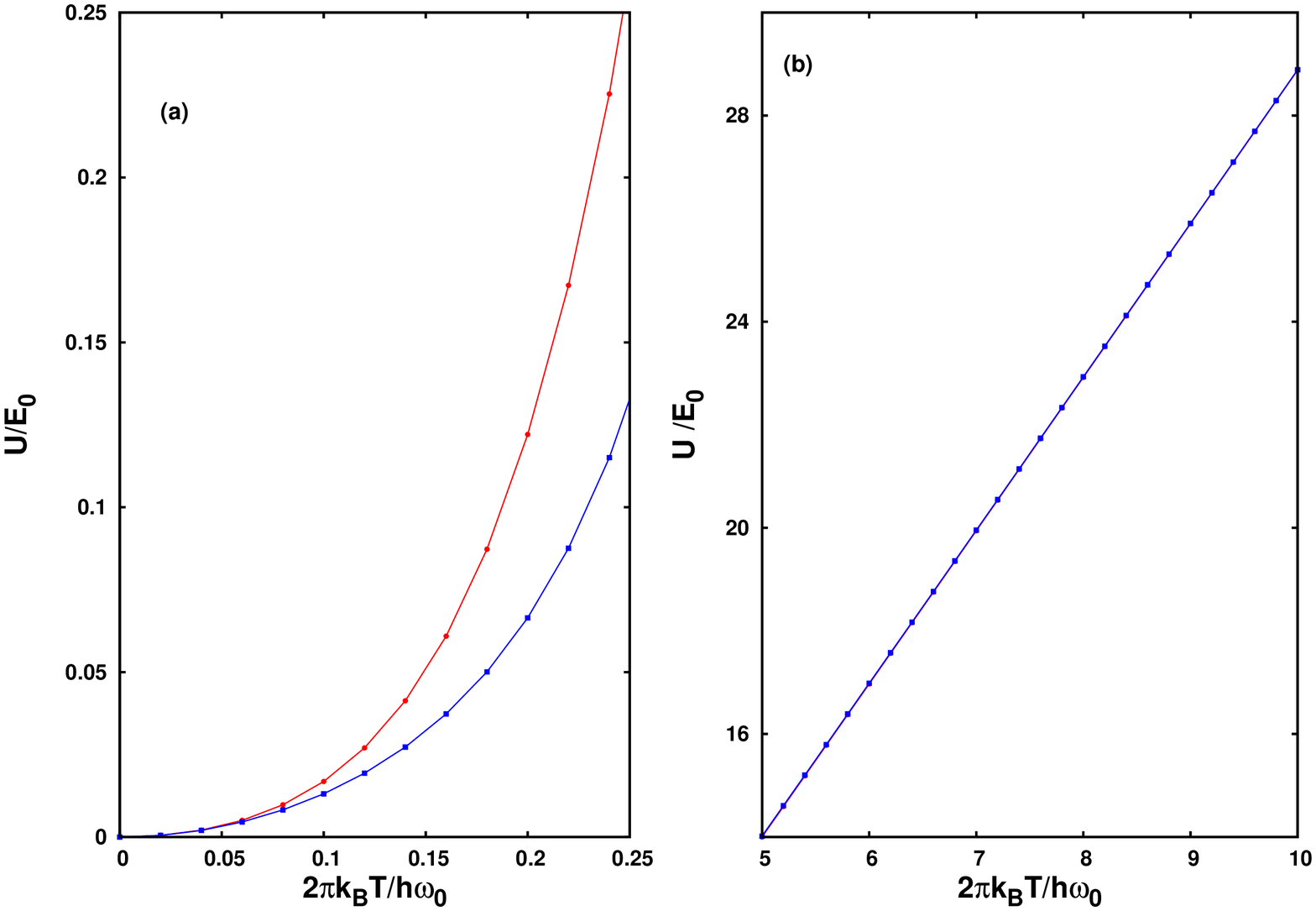}}}}
%\vskip-2.5cm
\caption{(color online)  Plot of $\frac{U}{E_{0}}$ versus dimensionless temperature, $\frac{2\pi k_BT}{h\omega_0}$, for the charged magneto-oscillator  (red filled circle) and for the free  charged oscillator (blue filled square) coupled to a Ohmic or SRT heat bath (a) in the low temperature regime and (b) in the high temperature regime. To plot this figure, we use $\frac{\gamma}{\omega_0}=0.8$, $\frac{\gamma}{\omega_c}=1.6$, and $\frac{\omega_0}{\omega_c}=2.0$.}
\end{center}
\end{figure}
\begin{figure}[h]
\begin{center}
{\rotatebox{0}{\resizebox{12cm}{6cm}{\includegraphics{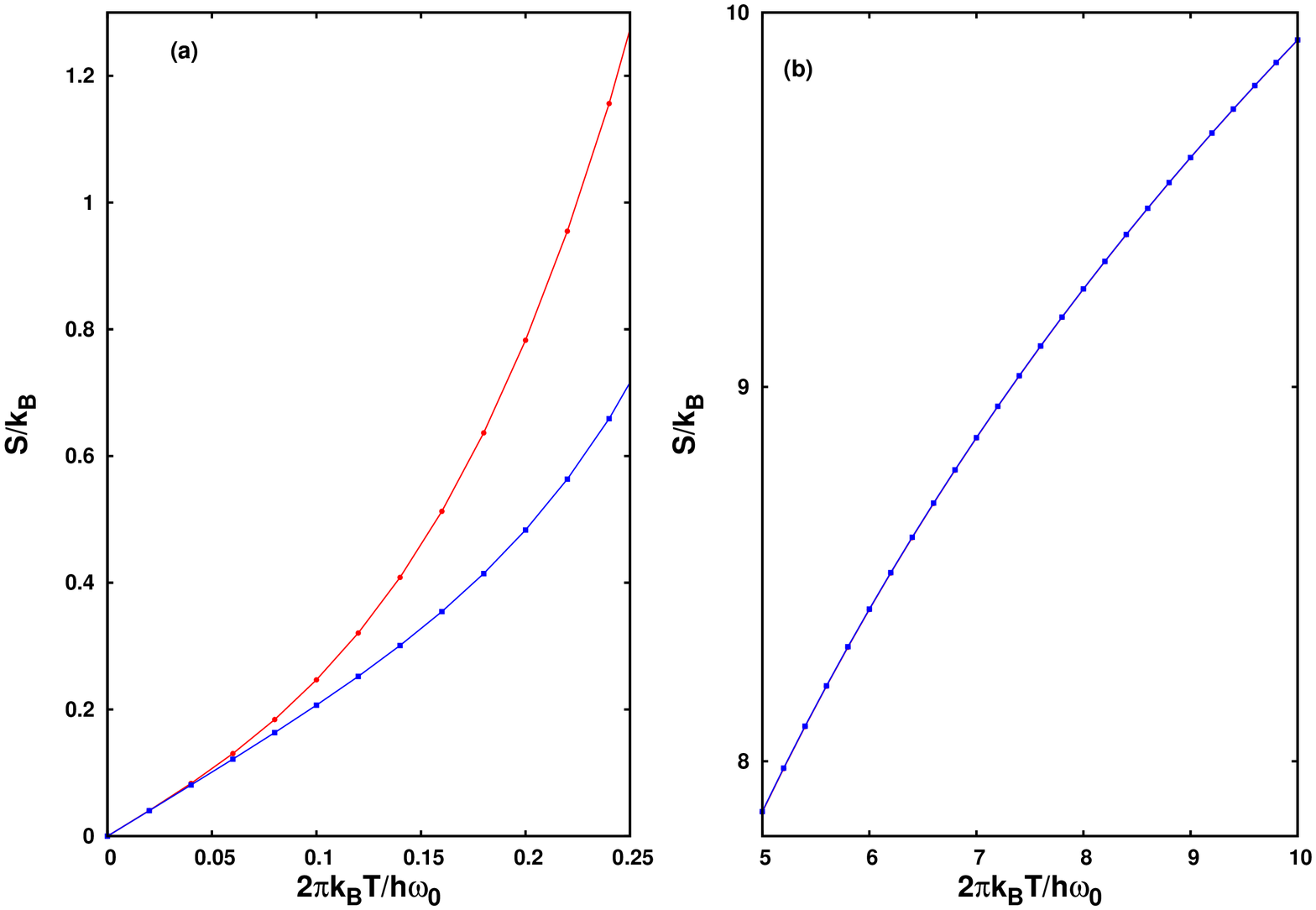}}}}
%\vskip-2.5cm
\caption{(color online) Plot of $\frac{S}{k_B}$ versus dimensionless temperature, $\frac{2\pi k_BT}{h\omega_0}$, for the charged magneto-oscillator  (red filled circle) and for the free  charged oscillator (blue filled square) coupled to a Ohmic or SRT heat bath (a) in the low temperature regime and (b) in the high temperature regime. To plot this figure, we use $\frac{\gamma}{\omega_0}=0.8$, $\frac{\gamma}{\omega_c}=1.6$, and $\frac{\omega_0}{\omega_c}=2.0$.}
\end{center}
\end{figure}
\begin{figure}[h]
\begin{center}
{\rotatebox{0}{\resizebox{12cm}{6cm}{\includegraphics{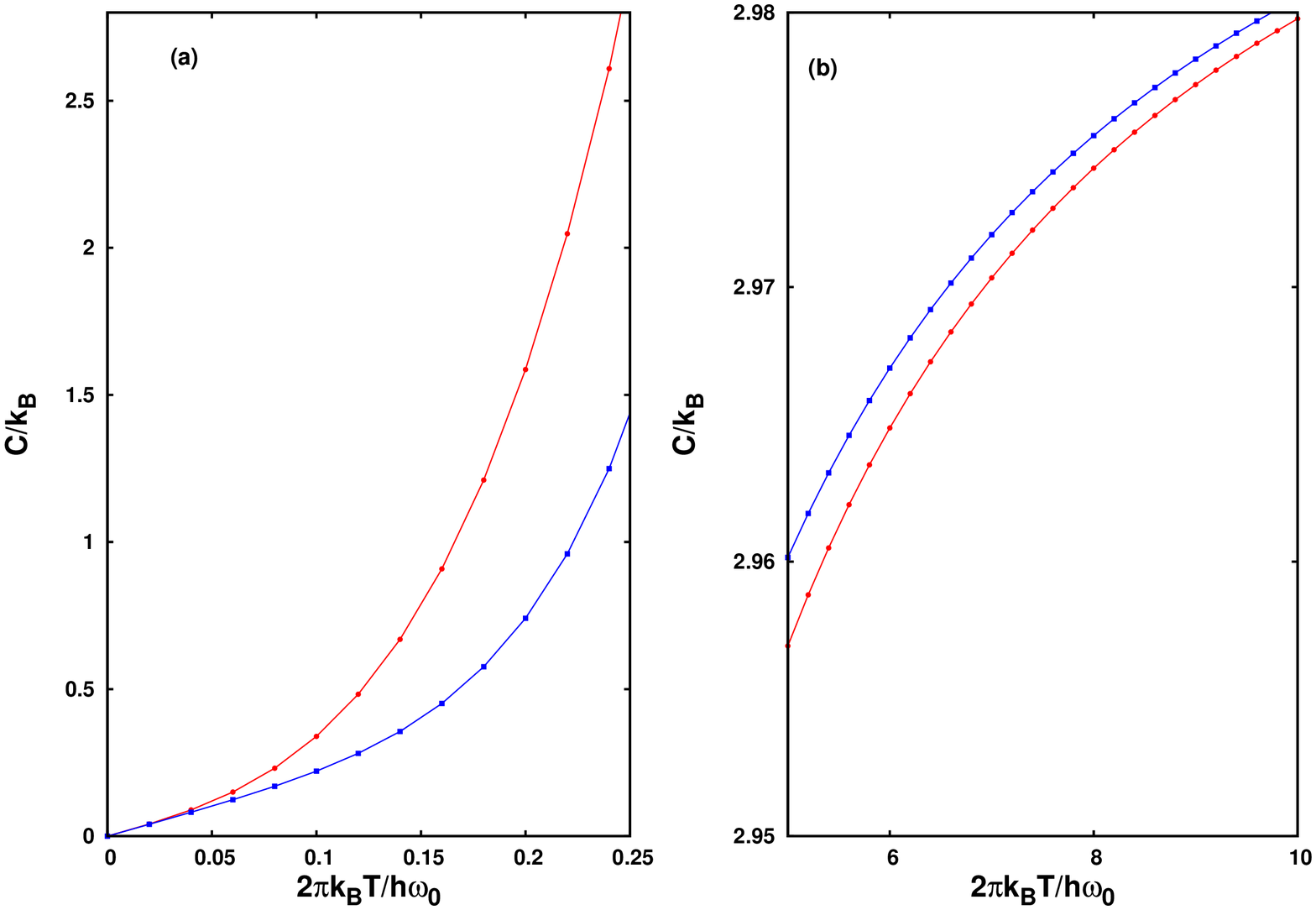}}}}
%\vskip-2.5cm
\caption{(color online) Plot of $\frac{C}{k_B}$ versus dimensionless temperature, $\frac{2\pi k_BT}{h\omega_0}$, for the charged magneto-oscillator (red filled circle) and for the free  charged oscillator (blue filled square) coupled to a Ohmic or SRT heat bath (a) in the low temperature regime and (b) in the high temperature regime. To plot this figure, we use $\frac{\gamma}{\omega_0}=0.8$, $\frac{\gamma}{\omega_c}=1.6$, and $\frac{\omega_0}{\omega_c}=2.0$. }
\end{center}
\end{figure} 
With the general form of $\alpha^{(0)}(\omega)$ (Eq. 22) one can write free energy as follows
\begin{eqnarray}
%\hskip-1.0cm
F(T,B)&=&\frac{3k_BT}{\pi}\int_0^{\infty}d\omega\ln\Big(1-e^{-\frac{\hbar\omega}{k_BT}}\Big)\Big(-\frac{\Omega}{\omega^2+\Omega^2}+\frac{\Omega^{\prime}}{\omega^2+{\Omega^{\prime}}^2}\Big)\nonumber \\
%\hskip-1.0cm
&&+\frac{k_BT}{\pi}\int_0^{\infty}d\omega\ln\Big(1-e^{-\frac{\hbar\omega}{k_BT}}\Big)\Big(\frac{\omega_1}{\omega^2+\omega_1^2}+\frac{\omega_1^*}{\omega^2+{\omega_1^*}^2}\nonumber \\
%\hskip-1.0cm
&&+\frac{\Omega_1}{\omega^2+\Omega_1^2}+\frac{\Omega_1^*}{\omega^2+{\Omega_1^*}^2}+\frac{\Omega_2}{\omega^2+{\Omega_2}^2}+\frac{\Omega_2^*}{\omega^2+{\Omega_2^*}^2}\Big),
\end{eqnarray}
where $\omega_1=\frac{\gamma}{2}+i\sqrt{\omega_0^2-\frac{\gamma^2}{4}}$, $\Omega_1=\Big\lbrack\frac{\gamma}{2}+\Big(\frac{b-a}{2}\Big)^{\frac{1}{2}}\Big\rbrack-i\Big\lbrack\frac{\omega_c}{2}+\Big(\frac{b+a}{2}\Big)^{\frac{1}{2}}\Big\rbrack$, $\Omega_2=\Big\lbrack\frac{\gamma}{2}-\Big(\frac{b-a}{2}\Big)^{\frac{1}{2}}\Big\rbrack-i\Big\lbrack\frac{\omega_c}{2}-\Big(\frac{b+a}{2}\Big)^{\frac{1}{2}}\Big\rbrack$, $a =\Big(\frac{\omega_c}{2}\Big)^2+\Big(\omega_0^2-\frac{\gamma^2}{4}\Big)$; and $b=\Big\lbrack a^2+\Big(\frac{\gamma\omega_c}{2}\Big)^2\Big\rbrack^{\frac{1}{2}}$. $\omega_1^*$, $\Omega_1^*$, and $\Omega_2^*$ are complex conjugate of $\omega_1$, $\Omega_1$ and $\Omega_2$ respectively and $\omega_c=eB/mc$ is the cyclotron frequency. 
\begin{figure}[h]
\begin{center}
{\rotatebox{0}{\resizebox{12cm}{8cm}{\includegraphics{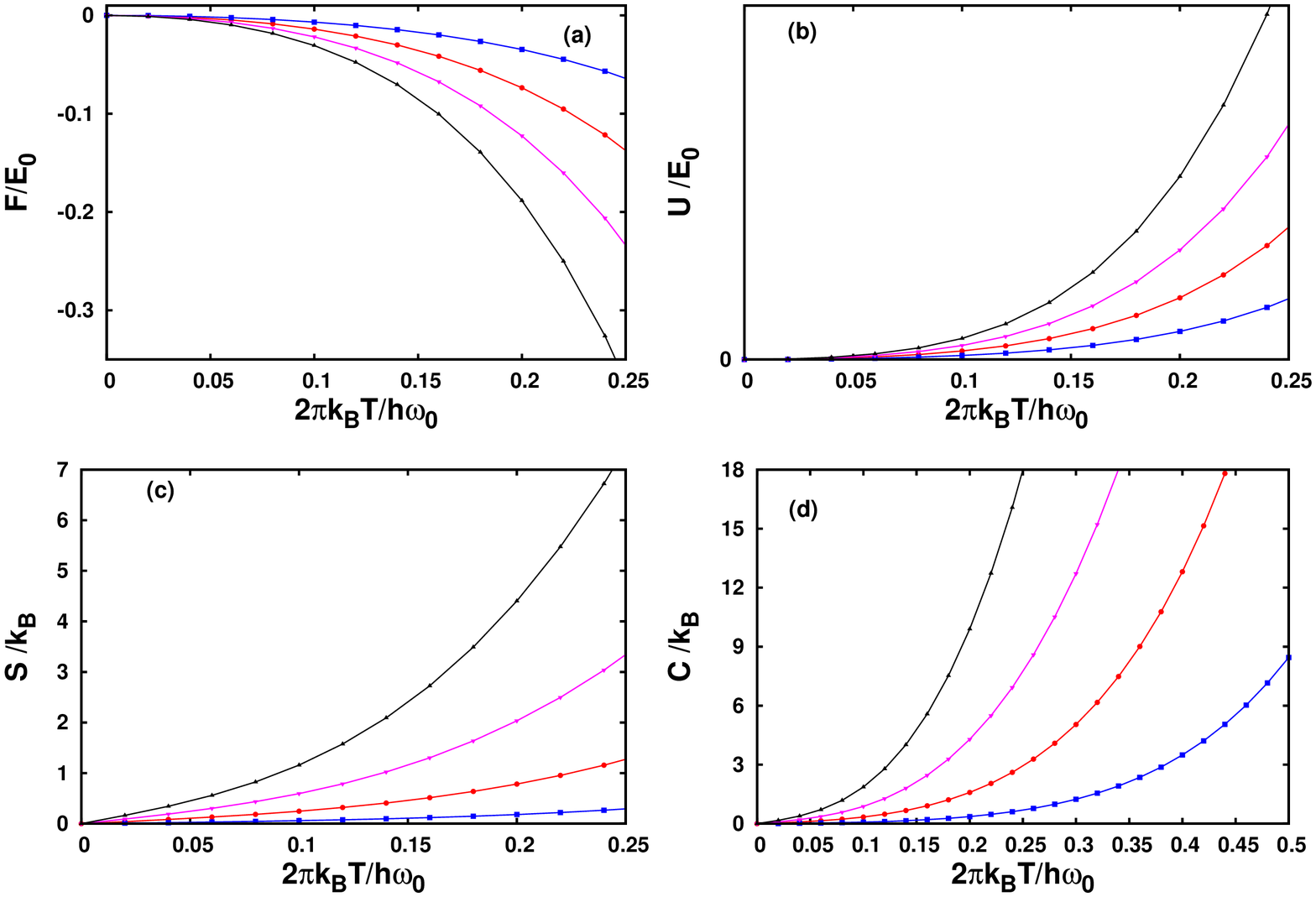}}}}
%\vskip-2.5cm
\caption{(color online) Plot of (a) $\frac{F}{E_{0}}$, (b)$\frac{U}{E_{0}}$, (c)$\frac{S}{k_{B}}$, and (d)$\frac{C}{k_{B}}$, versus dimensionless temperature, $\frac{2\pi k_BT}{h\omega_0}$, for the charged magneto-oscillator coupled to a Ohmic heat bath in the low temperature regime for different values of dissipation parameter, $\gamma$, blue filled square ($\gamma/\omega_0=0.25$), red filled circle ($\gamma/\omega_0=0.5$), pink upward triangle ($\gamma/\omega_0=0.75$) and black dowanward triangle ($\gamma/\omega_0=1.0$). To plot this figure, we also use $\frac{\omega_0}{\omega_c}=2.0$.}
\end{center}
\end{figure}
\begin{figure}[h]
\begin{center}
{\rotatebox{0}{\resizebox{12cm}{8cm}{\includegraphics{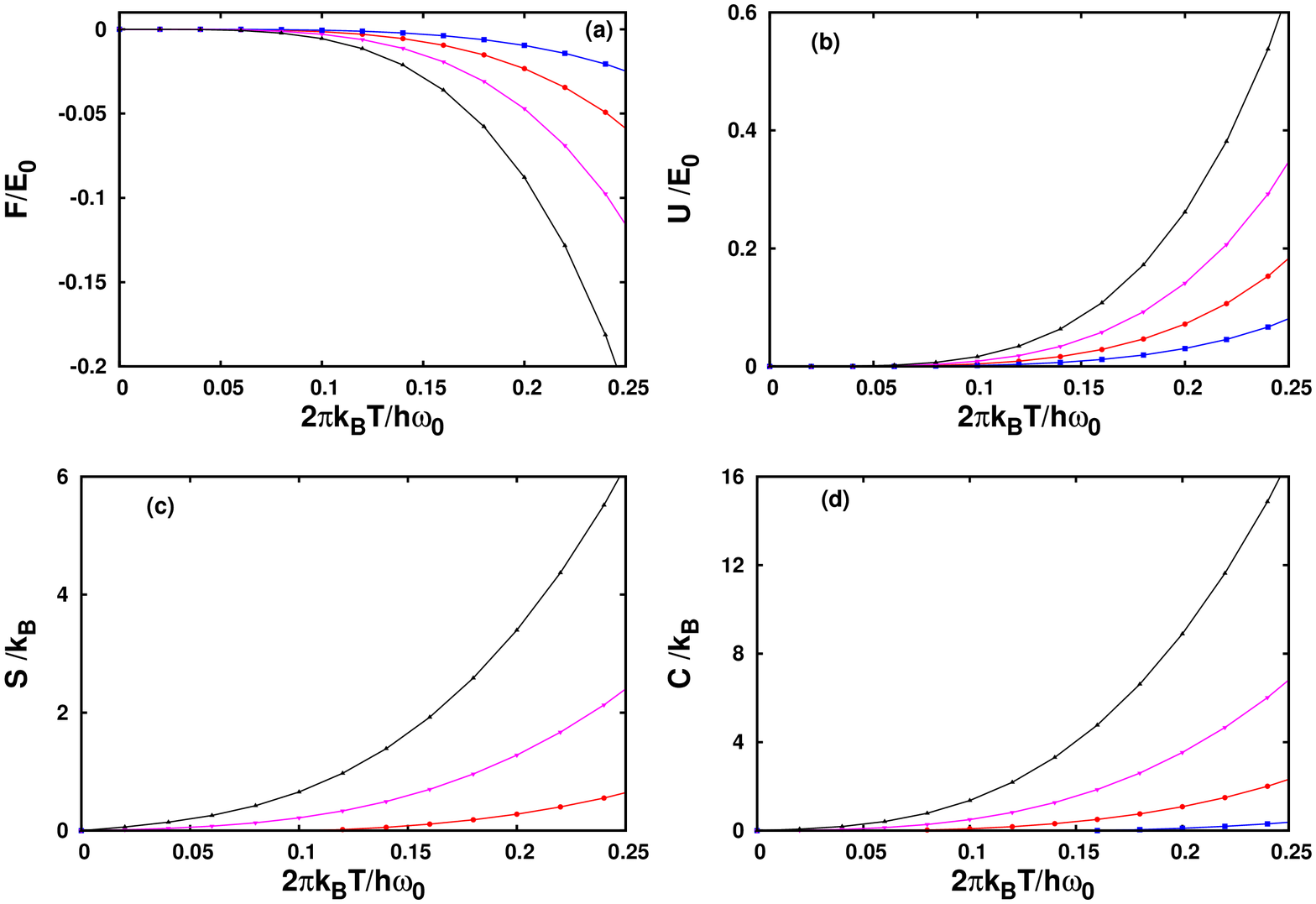}}}}
%\vskip-2.5cm
\caption{(color online) Plot of (a) $\frac{F}{E_{0}}$, (b)$\frac{U}{E_{0}}$, (c)$\frac{S}{k_{B}}$, and (d)$\frac{C}{k_{B}}$, versus dimensionless temperature, $\frac{2\pi k_BT}{h\omega_0}$, for the charged magneto-oscillator coupled to a radiation heat bath in the low temperature regime for different values of dissipation parameter, $\gamma$, blue filled square ($\gamma/\omega_0=0.25$), red filled circle ($\gamma/\omega_0=0.5$), pink upward triangle ($\gamma/\omega_0=0.75$) and black dowanward triangle ($\gamma/\omega_0=1.0$). To plot this figure, we also use $\frac{\omega_0}{\omega_c}=2.0$.}
\end{center}
\end{figure}
One can rewrite Eq. (23) in terms of Stieltjes function ($J(z)$) as follows :
\begin{eqnarray}
\hskip-1.0cm
&&F(T,B)=3k_BT\Big\lbrack J\Big(\frac{\hbar\Omega}{2\pi k_BT}\Big)-J\Big(\frac{\hbar\Omega^{\prime}}{2\pi k_BT}\Big)\Big\rbrack-k_BT\Big\lbrack J\Big(\frac{\hbar\omega_1}{2\pi k_BT}\Big)\nonumber \\
\hskip-1.0cm
&&+J\Big(\frac{\hbar\omega_1^*}{2\pi k_BT}\Big)+J\Big(\frac{\hbar\Omega_1}{2\pi k_BT}\Big)+J\Big(\frac{\hbar\Omega_1^*}{2\pi k_BT}\Big)+J\Big(\frac{\hbar\Omega_2}{2\pi k_BT}\Big)+J\Big(\frac{\hbar\Omega_2^*}{2\pi k_BT}\Big)\Big\rbrack,
\end{eqnarray}
where Stieltjes $J$ function is given by \cite{wall}
\begin{equation}
J(z)=-\frac{1}{\pi}\int_0^{\infty}dt\ln\Big(1-e^{-2\pi t}\Big)\frac{z}{t^2+z^2}.
\end{equation}
Now, we have all the ingredients to calculate the thermodynamic functions for the charged magneto-oscillator in contact with a heat bath. In the next section, we discuss about the Ohmic heat bath model.\\
\section{Ohmic Model}
In this section, we discuss the QTF of the charged magneto-oscillator in contact with a Ohmic heat bath in the low temperature as well as high temperature regime.
For the Ohmic model, $\Omega\rightarrow\infty$, $\tau\rightarrow 0$, $\frac{\gamma_0}{m}\rightarrow \gamma$, $\frac{K}{m}\rightarrow \omega_0^2$ and hence
\begin{eqnarray}
\hskip-1.0cm
F(T,B)&=&-k_BT\Big\lbrack J\Big(\frac{\hbar\omega_1}{2\pi k_BT}\Big)+J\Big(\frac{\hbar\omega_1^*}{2\pi k_BT}\Big)+J\Big(\frac{\hbar\Omega_1}{2\pi k_BT}\Big)\nonumber\\
\hskip-1.0cm
&&+J\Big(\frac{\hbar\Omega_1^*}{2\pi k_BT}\Big)+J\Big(\frac{\hbar\Omega_2}{2\pi k_BT}\Big)+J\Big(\frac{\hbar\Omega_2^*}{2\pi k_BT}\Big)\Big\rbrack. 
\end{eqnarray}
\subsection{Low-temperature expansion ($k_BT<<\hbar\omega_0$)}
In the low-temperature case, we use the asymptotic expansion for $J(z)$ :
\begin{equation}
J(z)=\sum_{n=0}^{\infty}\frac{D_{2n+2}}{(2n+1)(2n+2)}\frac{1}{z^{2n+1}},
\end{equation}
with $D_2=\frac{1}{6}$; $D_4=-\frac{1}{30}$; $D_6=\frac{1}{42}$; $D_8=-\frac{1}{30}$ and so on. 
\begin{figure}[h]
\begin{center}
{\rotatebox{0}{\resizebox{12cm}{8cm}{\includegraphics{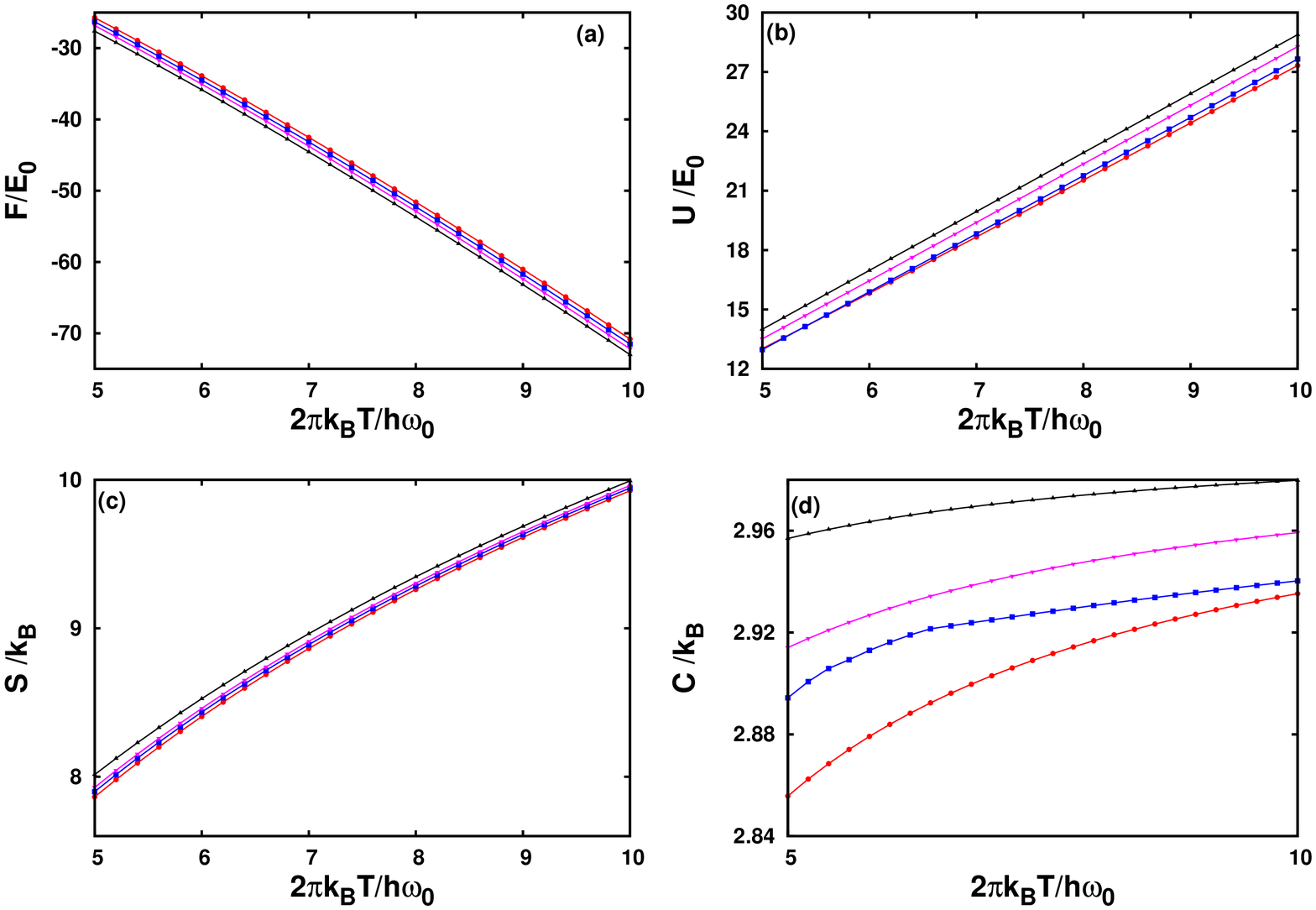}}}}
%\vskip-2.5cm
\caption{(color online) Plot of (a) $\frac{F}{E_{0}}$, (b)$\frac{U}{E_{0}}$, (c)$\frac{S}{k_{B}}$, and (d)$\frac{C}{k_{B}}$, versus dimensionless temperature, $\frac{2\pi k_BT}{h\omega_0}$, for the charged magneto-oscillator coupled to a Ohmic heat bath in the high temperature regime for different values of dissipation parameter, $\gamma$, red filled circle ($\gamma/\omega_0=0.25$), blue filled square ($\gamma/\omega_0=0.5$), pink upward triangle ($\gamma/\omega_0=0.75$) and black dowanward triangle ($\gamma/\omega_0=1.0$). To plot this figure, we also use $\frac{\omega_0}{\omega_c}=2.0$}
\end{center}
\end{figure}
\begin{figure}[h]
\begin{center}
{\rotatebox{0}{\resizebox{12cm}{8cm}{\includegraphics{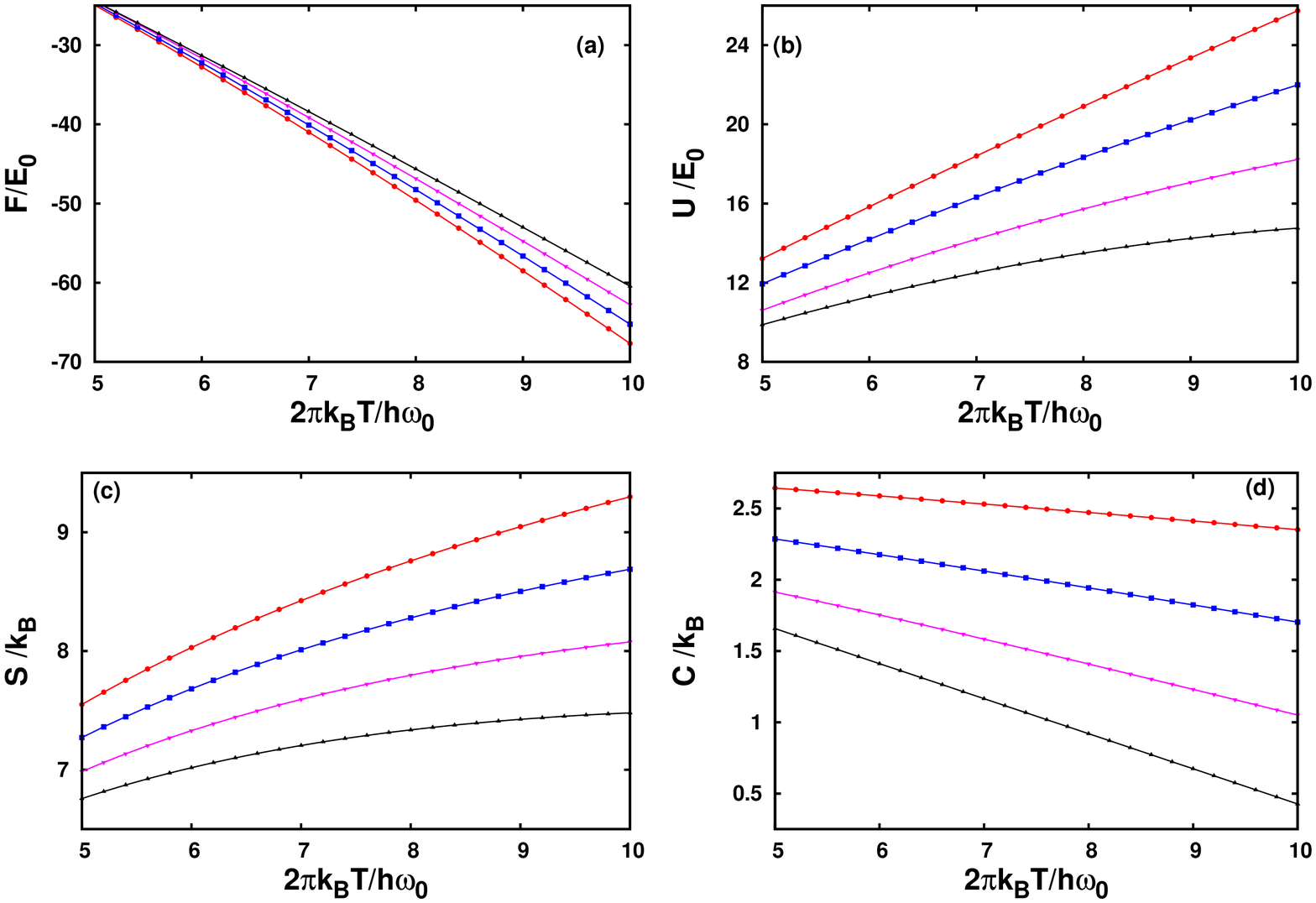}}}}
%\vskip-2.5cm
\caption{(color online) Plot of (a) $\frac{F}{E_{0}}$, (b)$\frac{U}{E_{0}}$, (c)$\frac{S}{k_{B}}$, and (d)$\frac{C}{k_{B}}$, versus dimensionless temperature, $\frac{2\pi k_BT}{h\omega_0}$, for the charged magneto-oscillator coupled to a radiation heat bath in the high temperature regime for different values of dissipation parameter, $\gamma$, red filled circle ($\gamma/\omega_0=0.25$), blue filled square ($\gamma/\omega_0=0.5$), pink upward triangle ($\gamma/\omega_0=0.75$) and black dowanward triangle ($\gamma/\omega_0=1.0$). To plot this figure, we also use $\frac{\omega_0}{\omega_c}=2.0$ }
\end{center}
\end{figure}
With this asymptotic expansion, the free energy becomes :
\begin{eqnarray}
\hskip-1.0cm
F(T,B)&=&-\Big\lbrack\frac{\pi(k_BT)^2\gamma}{2\hbar\omega_0^2}+\frac{\pi^3(k_BT)^4}{45\hbar^3\omega_0^6}(A_1+A_2+A_3)\nonumber \\
&&+\frac{8\pi^5(k_BT)^6(B_1+B_2+B_3)}{315\hbar^5\omega_0^{10}}+..\Big\rbrack,
\end{eqnarray}
where $A_1=\gamma(3\omega_0^2-\gamma^2)$, $A_2=(3\Gamma_1^2\lambda_1-\Lambda_1^3)$, $A_3=(3\Gamma_2^2\lambda_2-\Lambda_2^3)$, $B_1=\gamma(5\omega_0^4-5\gamma^2\omega_0^2+\gamma^4)$, $B_2=(\lambda_1^5-10\lambda_1^3\Gamma_1^2+5\lambda_1\Gamma_1^4)$, $B_3=(\lambda_2^5-10\lambda_2^3\Gamma_2^2+5\lambda_2\Gamma_2^4)$, $\Gamma_{1,2}=\frac{\omega_c}{2}\pm\sqrt{\frac{b+a}{2}}$; $\lambda_{1,2}=\frac{\gamma}{2}\pm\sqrt{\frac{b+a}{2}}$; and $\Lambda_{1,2}=\gamma\pm \sqrt{2(b-a)}$. Entropy can be written as 
\begin{eqnarray}
\hskip -0.4cm
S(T,B)&=&k_B\Big\lbrack\frac{\pi k_BT\gamma}{\hbar\omega_0^2}+\frac{4\pi^3(k_BT)^3}{45\hbar^3\omega_0^6}(A_1+A_2+A_3)\nonumber \\
\hskip-0.4cm
&&+\frac{16\pi^5(k_BT)^5(B_1+B_2+B_3)}{105\hbar^5\omega_0^{10}}+...\Big\rbrack.
\end{eqnarray}
\begin{figure}[h]
\begin{center}
{\rotatebox{0}{\resizebox{12cm}{8cm}{\includegraphics{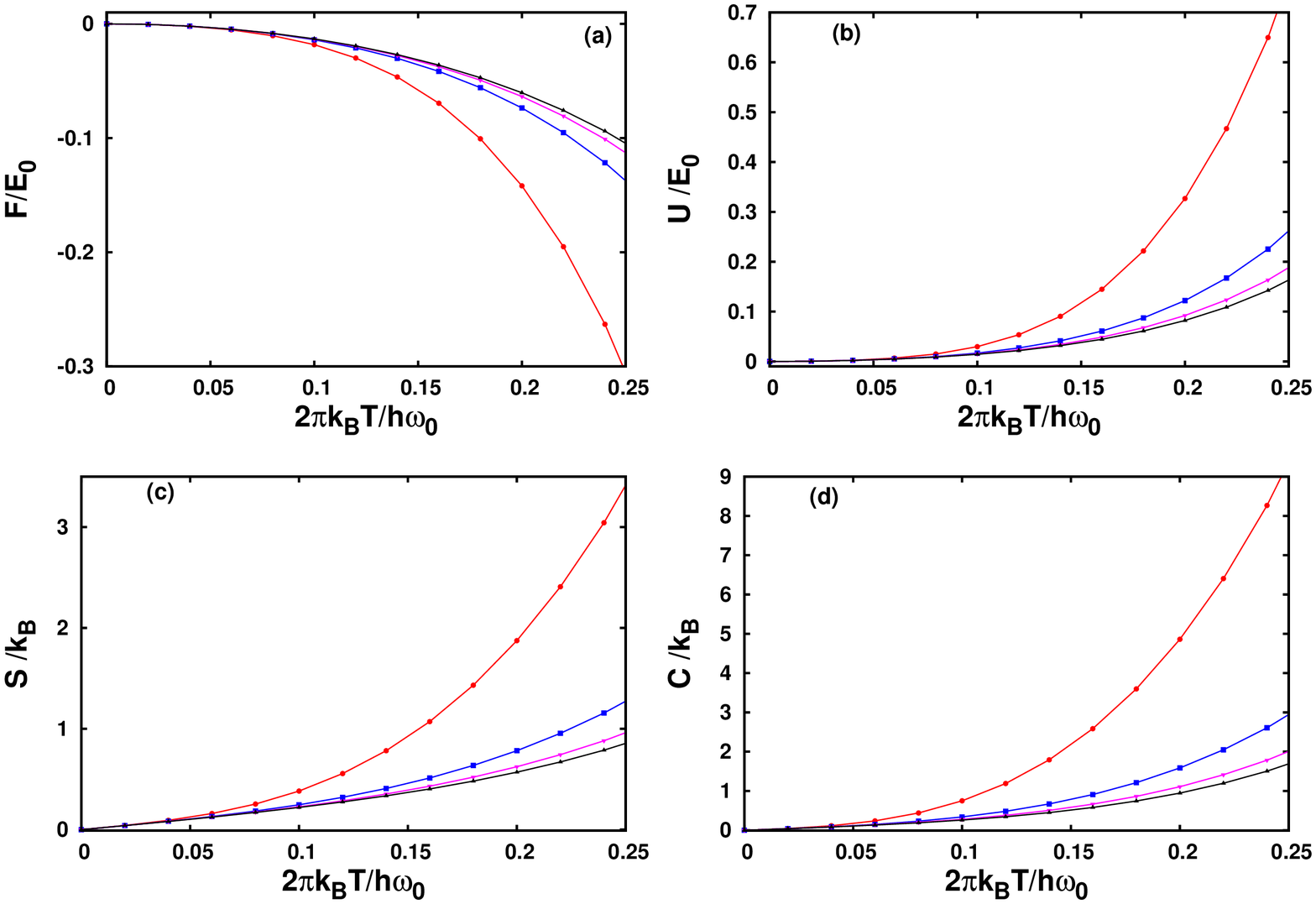}}}}
%\vskip-2.5cm
\caption{(color online) Plot of (a) $\frac{F}{E_{0}}$, (b)$\frac{U}{E_{0}}$, (c)$\frac{S}{k_{B}}$, and (d)$\frac{C}{k_{B}}$, versus dimensionless temperature, $\frac{2\pi k_BT}{h\omega_0}$, for the charged magneto-oscillator coupled to a Ohmic heat bath in the low temperature regime for different values of cyclotron frequency, $\omega_c$, red filled circle ($\omega_c/\omega_0=0.25$), blue filled square ($\omega_c/\omega_0=0.5$), pink upward triangle ($\omega_c/\omega_0=0.75$) and black dowanward triangle ($\omega_c/\omega_0=1.0$). To plot this figure, we also use $\frac{\gamma}{\omega_0}=0.8$. }
\end{center}
\end{figure}
\begin{figure}[h]
\begin{center}
{\rotatebox{0}{\resizebox{12cm}{8cm}{\includegraphics{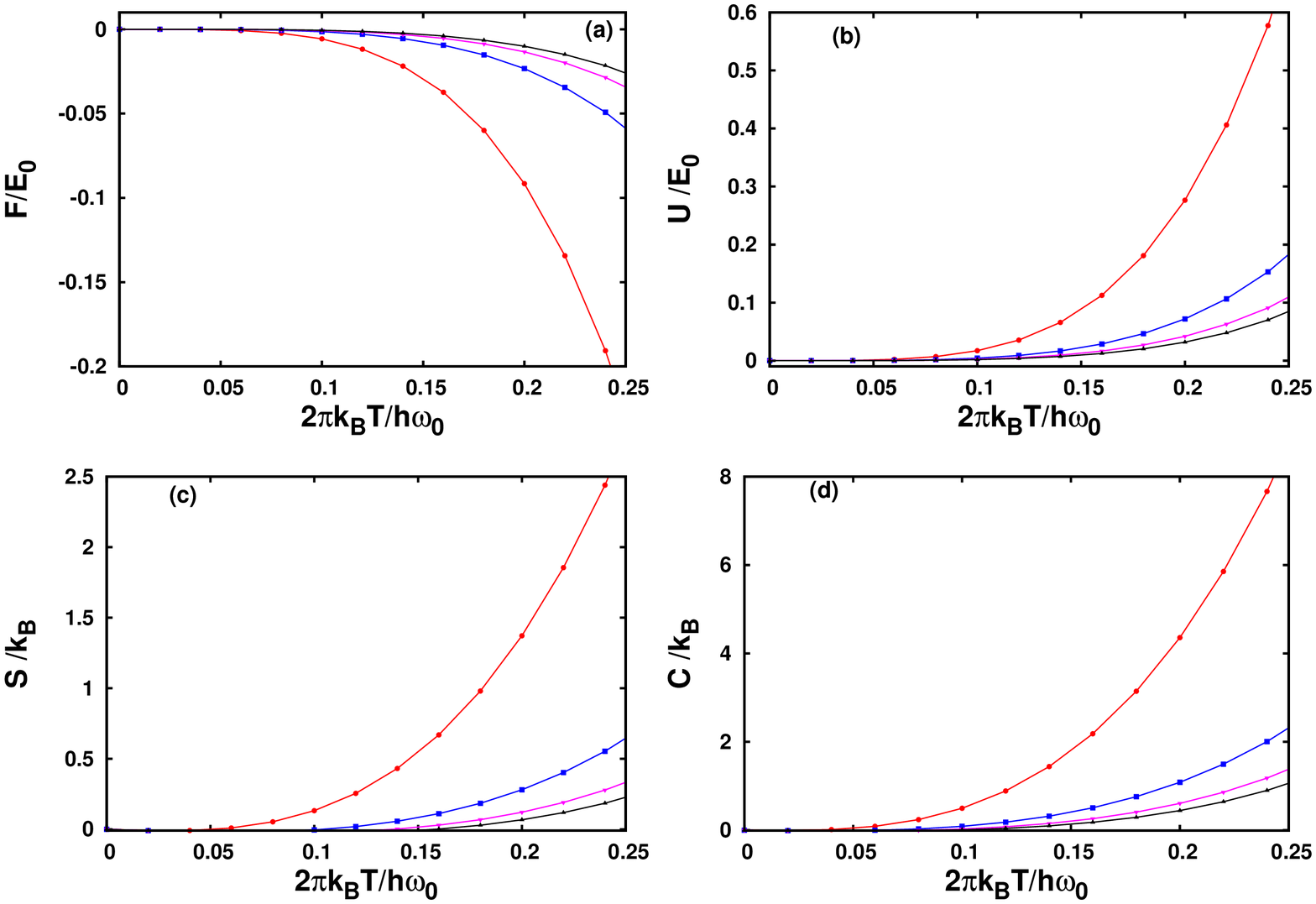}}}}
%\vskip-2.5cm
\caption{(color online) Plot of (a) $\frac{F}{E_{0}}$, (b)$\frac{U}{E_{0}}$, (c)$\frac{S}{k_{B}}$, and (d)$\frac{C}{k_{B}}$, versus dimensionless temperature, $\frac{2\pi k_BT}{h\omega_0}$, for the charged magneto-oscillator coupled to a radiation heat bath in the low temperature regime for different values of cyclotron frequency, $\omega_c$, red filled circle ($\omega_c/\omega_0=0.25$), blue filled square ($\omega_c/\omega_0=0.5$), pink upward triangle ($\omega_c/\omega_0=0.75$) and black dowanward triangle ($\omega_c/\omega_0=1.0$). To plot this figure, we also use $\frac{\gamma}{\omega_0}=0.8$. }
\end{center}
\end{figure}
Internal energy can be written as follows :
\begin{eqnarray}
\hskip-1.0cm
U(T,B)&=&\frac{\pi(k_BT)^2\gamma}{2\hbar\omega_0^2}+\frac{\pi^3(k_BT)^4(A_1+A_2+A_3)}{45\hbar^3\omega_0^6}\nonumber \\
&&+\frac{40\pi^5(k_BT)^5(B_1+B_2+B_3)}{315\hbar^5\omega_0^{10}}+...
\end{eqnarray}
Finally, specific heat of the system is given by
\begin{eqnarray}
\hskip-0.4cm
C(T,B)&=&k_B\Big\lbrack\frac{\pi k_BT\gamma}{\hbar\omega_0^2}+\frac{4\pi^3(k_BT)^3}{15\hbar^3\omega_0^6}(A_1+A_2+A_3)\nonumber \\
\hskip-0.4cm
&& +\frac{16\pi^5(k_BT)^5(B_1+B_2+B_3)}{21\hbar^5\omega_0^{10}}+...\Big\rbrack
\end{eqnarray}
\subsection{High-temperature expansion ($k_BT>>\hbar\omega_0$)}
In the high temperature regime, we use the small arguement expansion for $J(z)$ :
\begin{eqnarray}
J(z)&=&-\frac{1}{2}\ln(2\pi)-(z+\frac{1}{2})\ln z+(1-\gamma_E)z+\sum_{n=2}^{\infty}\frac{(-1)^n\zeta(n)}{n}z^n,
\end{eqnarray}
where $\gamma_E=0.577215$ is Euler's constant and $\zeta(n)$ is the Riemann Zeta function. Using Eq. (32), one can write down the free energy of the system in the high temperature regime as follows:
\begin{eqnarray}
\hskip-1.0cm
&&F(T,B)=-3k_BT\ln\Big(\frac{k_BT}{\hbar\omega_0}\Big)-\frac{3\hbar\gamma}{2\pi}\ln\Big(\frac{2\pi k_BT}{\hbar\omega_0}\Big)-\frac{\hbar\omega_1^{\prime}}{\pi}\phi-\frac{3\hbar\gamma(1-\gamma_E)}{2\pi}\nonumber \\
\hskip-1.0cm
&&-\frac{2\hbar\omega_c}{\pi}\theta-2k_BT\Big\lbrack\sum_{n=2}^{\infty}(-1)^n\frac{\zeta(n)}{n}\Big(\frac{\hbar\omega_0}{2\pi k_BT}\Big)^n\Big(\tan(n\phi)+2\tan(n\theta)\Big)\Big\rbrack,
\end{eqnarray}
where $\theta = \tan^{-1}P$, $P=\frac{\sqrt{2(b+a)}}{\gamma}$, $\phi=\tan^{-1}(\frac{2\omega_1^{\prime}}{\gamma})$, and $\omega_1^{\prime}=\sqrt{\omega_0^2-\frac{\gamma^2}{4}}$. The entropy is given by
\begin{eqnarray}
\hskip-1.0cm
&&S(T,B)=3k_B\Big(\ln\frac{k_BT}{\hbar\omega_0}+1\Big)-4k_B\sum_{n=2}^{\infty}(-1)^n\frac{(n-1)\zeta(n)}{n}\Big(\frac{\hbar\omega_0}{2\pi k_BT}\Big)^n\tan(n\theta)\nonumber \\
\hskip-1.0cm
&&+\frac{3\hbar\gamma}{2\pi T}-2k_B\sum_{n=2}^{\infty}(-1)^n\frac{(n-1)\zeta(n)}{n}\Big(\frac{\hbar\omega_0}{2\pi k_BT}\Big)^n\tan(n\phi).
\end{eqnarray}
\begin{figure}[h]
\begin{center}
{\rotatebox{0}{\resizebox{12cm}{8cm}{\includegraphics{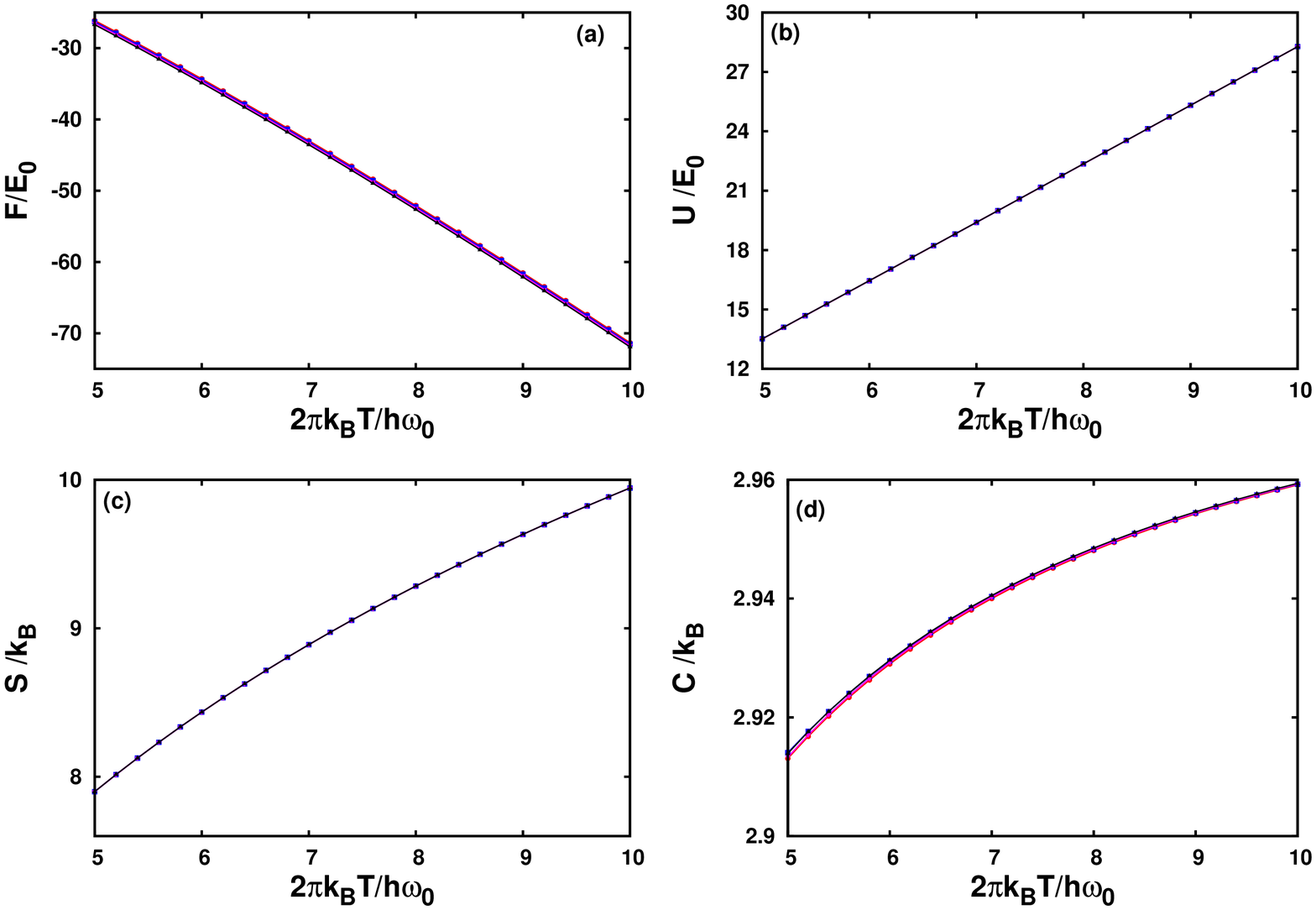}}}}
%\vskip-2.5cm
\caption{(color online) Plot of (a) $\frac{F}{E_{0}}$, (b)$\frac{U}{E_{0}}$, (c)$\frac{S}{k_{B}}$, and (d)$\frac{C}{k_{B}}$, versus dimensionless temperature, $\frac{2\pi k_BT}{h\omega_0}$, for the charged magneto-oscillator coupled to a Ohmic heat bath in the high temperature regime for different values of cyclotron frequency, $\omega_c$, red filled circle ($\omega_c/\omega_0=0.25$), blue filled square ($\omega_c/\omega_0=0.5$), pink upward triangle ($\omega_c/\omega_0=0.75$) and black dowanward triangle ($\omega_c/\omega_0=1.0$). To plot this figure, we also use $\frac{\gamma}{\omega_0}=0.8$. }
\end{center}
\end{figure}
\begin{figure}[h]
\begin{center}
{\rotatebox{0}{\resizebox{12cm}{7cm}{\includegraphics{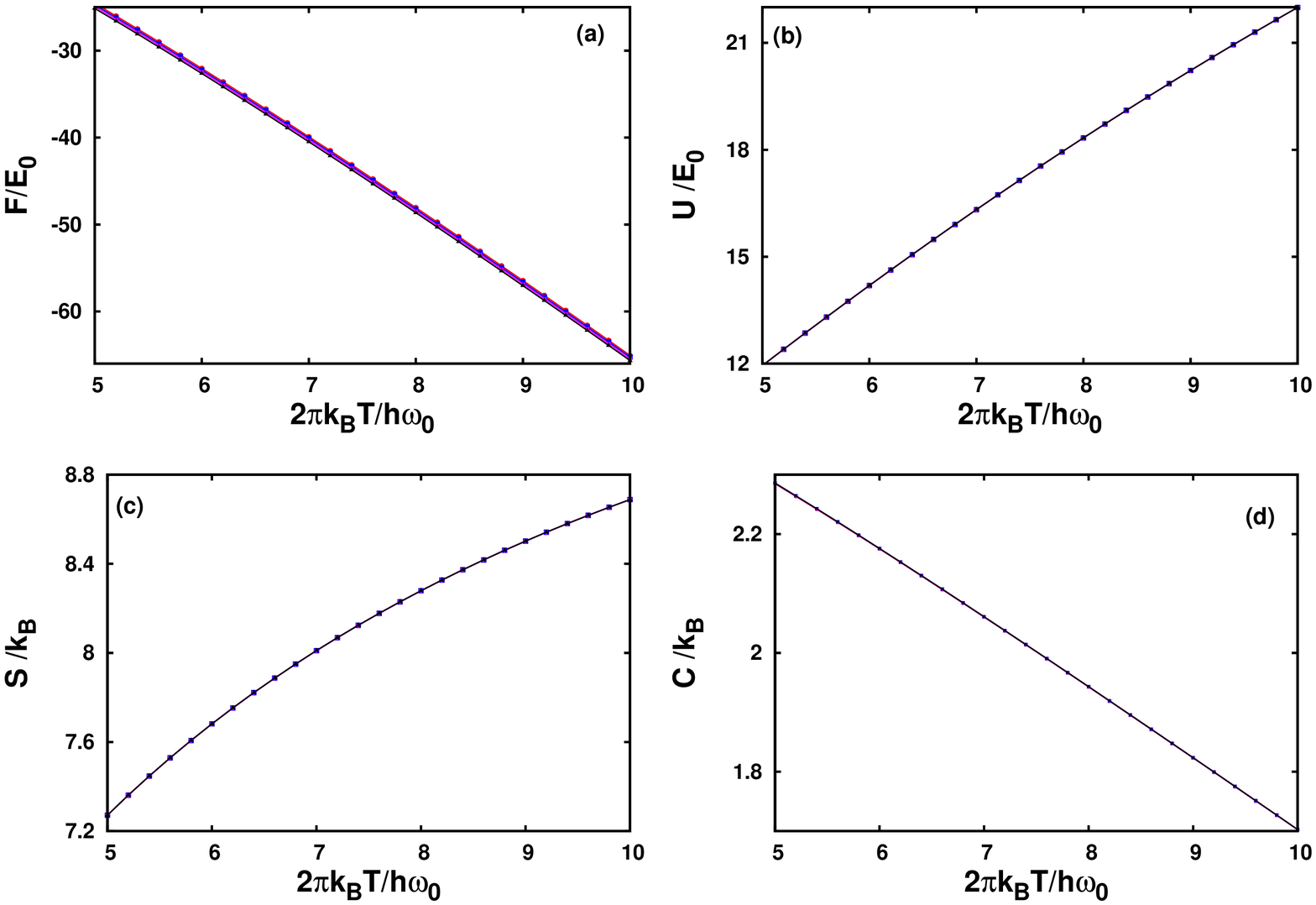}}}}
%\vskip-2.5cm
\caption{(color online) Plot of (a) $\frac{F}{E_{0}}$, (b)$\frac{U}{E_{0}}$, (c)$\frac{S}{k_{B}}$, and (d)$\frac{C}{k_{B}}$, versus dimensionless temperature, $\frac{2\pi k_BT}{h\omega_0}$, for the charged magneto-oscillator coupled to a radiation heat bath in the high temperature regime for different values of cyclotron frequency, $\omega_c$, red filled circle ($\omega_c/\omega_0=0.25$), blue filled square ($\omega_c/\omega_0=0.5$), pink upward triangle ($\omega_c/\omega_0=0.75$) and black dowanward triangle ($\omega_c/\omega_0=1.0$). To plot this figure, we also use $\frac{\gamma}{\omega_0}=0.8$. }
\end{center}
\end{figure}
On the other hand, internal energy is given by
\begin{eqnarray}
\hskip-1.8cm
U(T,B)&=&3k_BT-\frac{3\hbar\gamma}{2\pi}\ln\Big(\frac{2\pi k_BT}{\hbar\omega_0}\Big) -\frac{\hbar\omega_1^{\prime}}{\pi}\phi+\frac{3\hbar\gamma\gamma_E}{2\pi}-\frac{2\hbar\omega_c}{\pi}\theta \nonumber \\
&&-4k_BT\sum_{n=2}^{\infty}(-1)^n\frac{(n-1)\zeta(n)}{n}\Big(\frac{\hbar\omega_0}{2\pi k_BT}\Big)^n\tan(n\phi)\nonumber \\
&& -8k_BT\sum_{n=2}^{\infty}(-1)^n\frac{\zeta(n)}{n}\Big(\frac{\hbar\omega_0}{2\pi k_BT}\Big)^n\tan(n\theta).
\end{eqnarray}
Finally, specific heat is given by
\begin{eqnarray}
\hskip -1.0cm
C(T,B)&=&3k_B-\frac{3\hbar\gamma}{2\pi T}+4k_B\sum_{n=2}^{\infty}(-1)^n(n-1)\zeta(n)\Big(\frac{\hbar\omega_0}{2\pi k_BT}\Big)^n\tan(n\theta)\nonumber \\
\hskip-1.0cm
&&+2k_B\sum_{n=2}^{\infty}(-1)^n(n-1)\zeta(n)\Big(\frac{\hbar\omega_0}{2\pi k_BT}\Big)^n\tan(n\phi).
\end{eqnarray}
\section{Single relaxation time and QED model}
Free energy for the single relaxation time and QED model is given by
\begin{equation}
F(T,B)=F_{ohmic}(T,B)+3k_BT\Big\lbrack J\Big(\frac{\hbar\Omega}{2\pi k_BT}\Big)-J\Big(\frac{\hbar\Omega^{\prime}}{2\pi k_BT}\Big)\Big\rbrack.
\end{equation}
As $\Omega$ and $\Omega^{\prime}$ are always large compared with $k_BT$, so retaining only the first term of the low temperature expansion of $J(z)$; one can write 
\begin{equation}
F(T,B)=F_{ohmic}(T,B)+\frac{\pi(k_BT)^2}{2\hbar}\Big(\frac{1}{\Omega}-\frac{1}{\Omega^{\prime}}\Big).
\end{equation}
For single relaxation time model, $\frac{1}{\Omega}=\frac{1}{\Omega^{\prime}+\gamma}$ and $\Omega^{\prime}>>\gamma$. Hence, the second term in Eq. (38) is negligibly small and 
\begin{equation}
F_{SRT}(T,B)\simeq F_{Ohmic}(T,B),
\end{equation}
where subscript SRT stands for single relaxation time. On the other hand, for QED, $\frac{1}{\Omega}-\frac{1}{\Omega^{\prime}}=\frac{\gamma}{\omega_0^2}$. Thus, 
\begin{equation}
F_{QED}(T,B)=F_{Ohmic}(T,B)+\frac{\pi(k_BT)^2\gamma}{2\hbar\omega_0^2}.
\end{equation}
\subsection{QED : Low temperature expansion ($k_BT<<\hbar\omega_0$)}
Using the low temperature expansion of $J(z)$, one can obtain
\begin{eqnarray}
\hskip-1.0cm
F_{QED}(T,B)&=&-\Big\lbrack \frac{\pi^3(k_BT)^4(A_1+A_2+A_3)}{45\hbar^3\omega_0^6}\nonumber \\
\hskip-1.0cm
&+&\frac{8\pi^5(k_BT)^6(B_1+B_2+B_3)}{315\hbar^5\omega_0^{10}}+....\rbrack.
\end{eqnarray}
The leading term in the expansion of $F_{Ohmic}(T,B)$ exactly cancels the second term in Eq. (40) and hence, the leading term is the $O(T^4)$. Entropy of the system is given by
\begin{eqnarray}
S_{QED}(T,B)&=&k_B\Big\lbrack\frac{4\pi^3(k_BT)^3(A_1+A_2+A_3)}{45\hbar^3\omega_0^6}\nonumber \\
&&+\frac{16\pi^5(k_BT)^5(B_1+B_2+B_3)}{105\hbar^5\omega_0^{10}}+...\rbrack
\end{eqnarray}
Internal energy is given by 
\begin{eqnarray}
U_{QED}(T,B)&=&\frac{\pi^3(k_BT)^4(A_1+A_2+A_3)}{15\hbar^3\omega_0^6}\nonumber \\
&&+\frac{8\pi^5(k_BT)^6(B_1+B_2+B_3)}{63\hbar^5\omega_0^{10}}+...
\end{eqnarray}
\begin{figure}[h]
\begin{center}
{\rotatebox{0}{\resizebox{12cm}{6cm}{\includegraphics{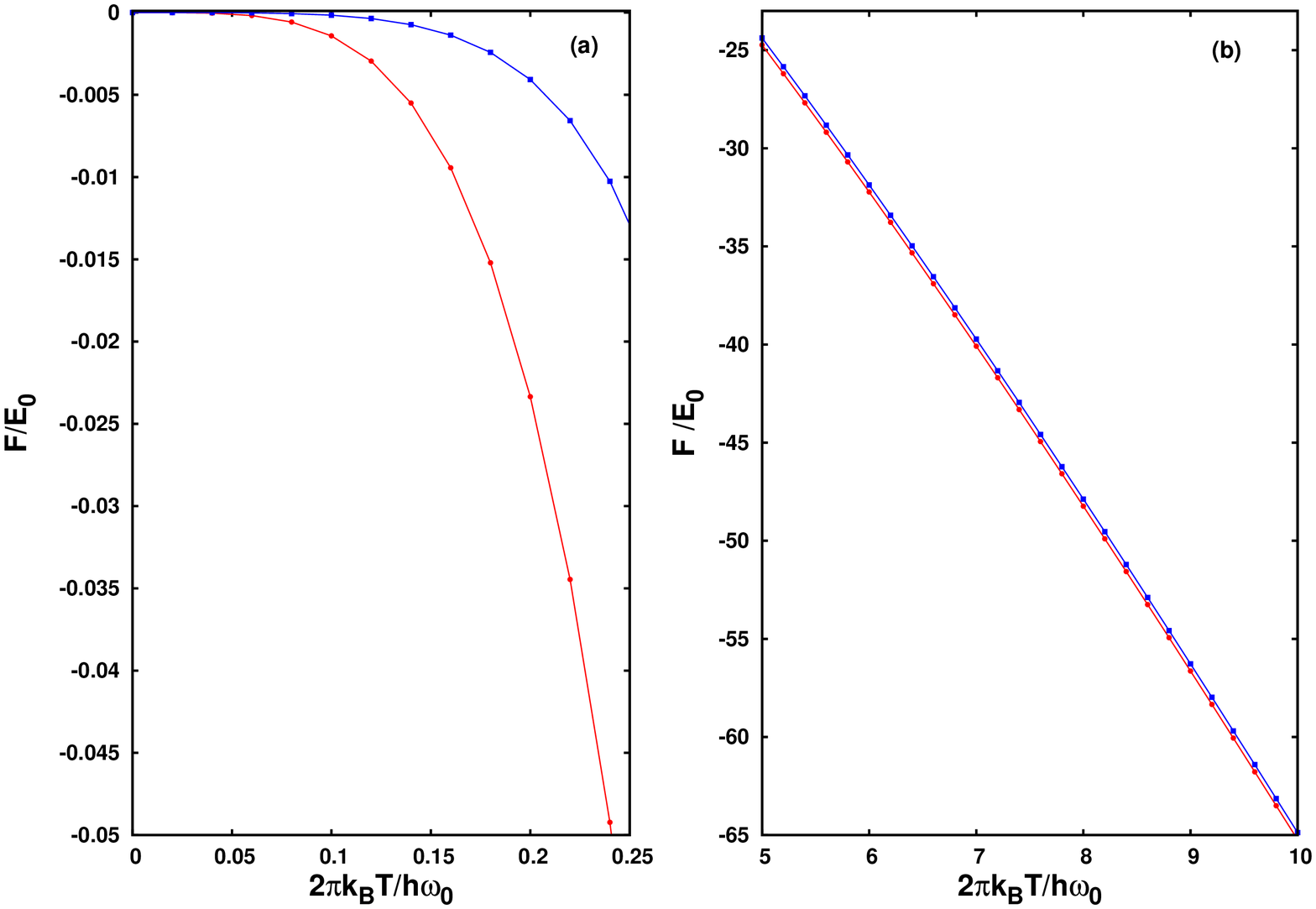}}}}
%\vskip-2.5cm
\caption{(color online) Plot of $\frac{F}{E_{0}}$ versus dimensionless temperature, $\frac{2\pi k_BT}{h\omega_0}$, for the charged magneto-oscillator (red filled circle) and for the free  charged oscillator (blue filled square) coupled to a blackbody radiation heat bath (a) in the low temperature regime and (b) in the high temperature regime. To plot this figure, we use $\frac{\gamma}{\omega_0}=0.8$, $\frac{\gamma}{\omega_c}=1.6$, and $\frac{\omega_0}{\omega_c}=2.0$.}
\end{center}
\end{figure}
\begin{figure}[h!]
\begin{center}
{\rotatebox{0}{\resizebox{12cm}{6cm}{\includegraphics{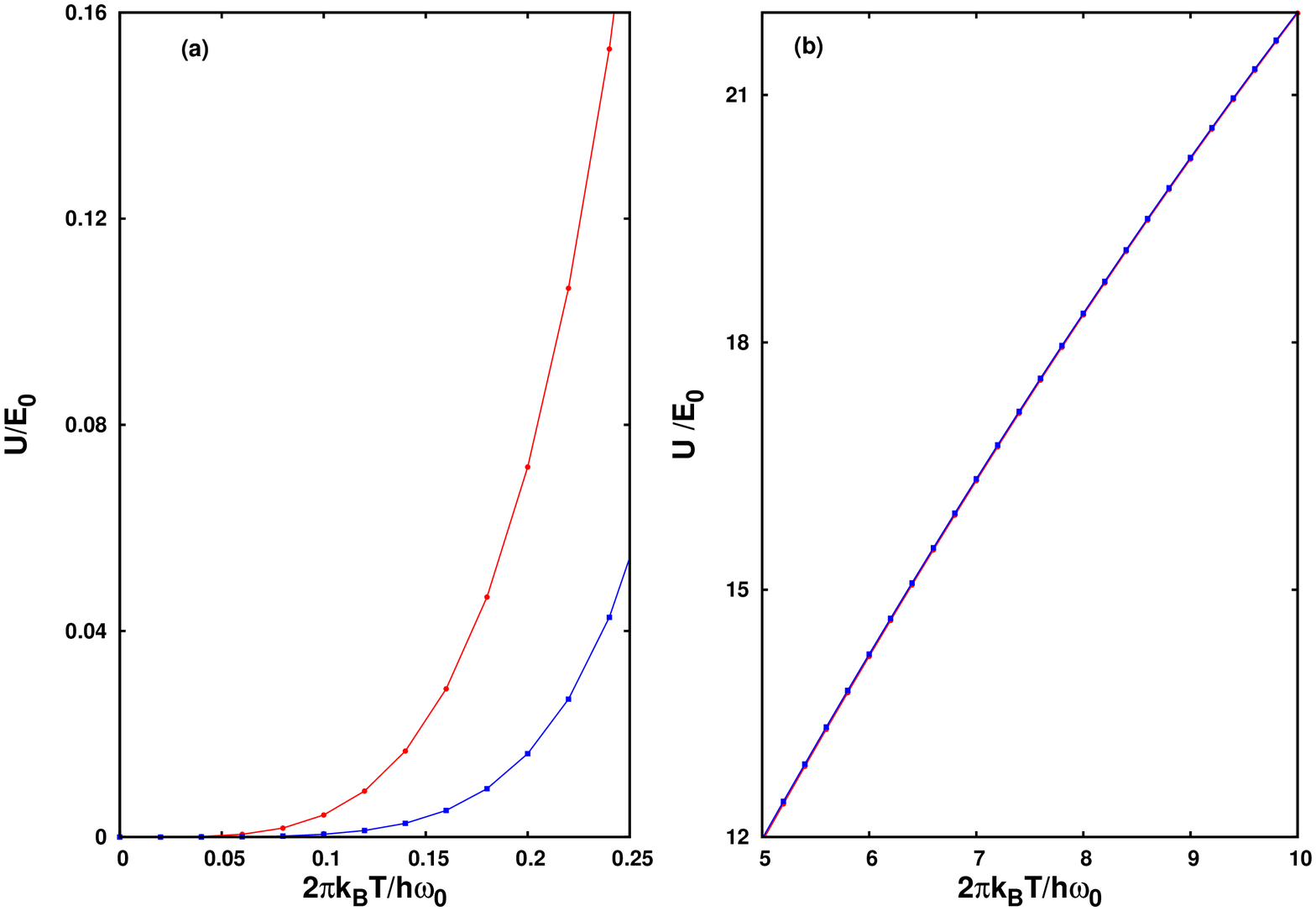}}}}
%\vskip-2.5cm
\caption{(color online)  Plot of $\frac{U}{E_{0}}$ versus dimensionless temperature, $\frac{2\pi k_BT}{h\omega_0}$, for the charged magneto-oscillator (red filled circle) and for the free  charged oscillator (blue filled square) coupled to a blackbody radiation heat bath (a) in the low temperature regime and (b) in the high temperature regime. To plot this figure, we use $\frac{\gamma}{\omega_0}=0.8$, $\frac{\gamma}{\omega_c}=1.6$, and $\frac{\omega_0}{\omega_c}=2.0$.}
\end{center}
\end{figure}  
\begin{figure}[h]
\begin{center}
{\rotatebox{0}{\resizebox{12cm}{6cm}{\includegraphics{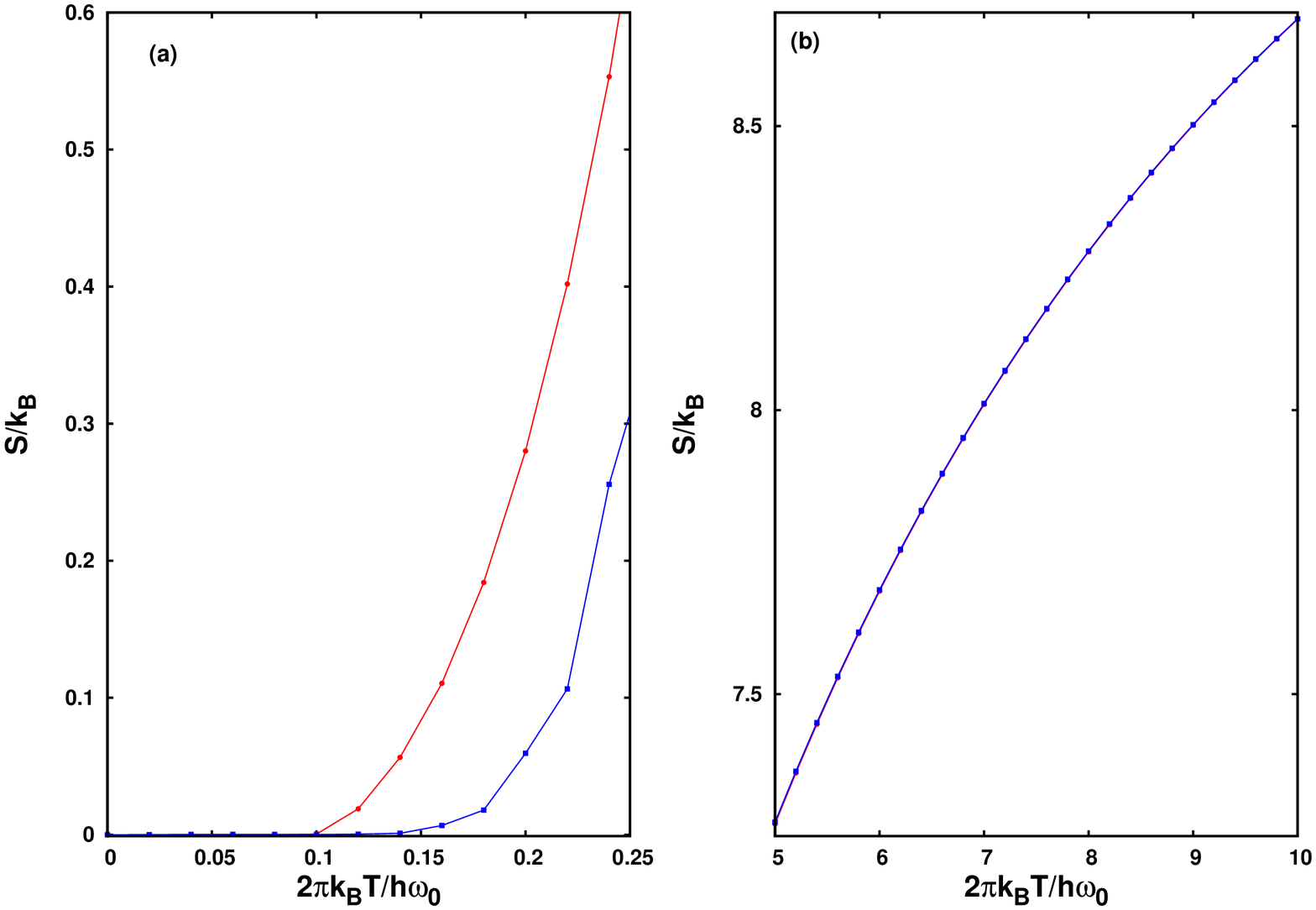}}}}
%\vskip-2.5cm
\caption{(color online)  Plot of $\frac{S}{k_B}$ versus dimensionless temperature, $\frac{2\pi k_BT}{h\omega_0}$, for the charged magneto-oscillator (red filled circle) and for the free  charged oscillator (blue filled square) coupled to a blackbody radiation heat bath (a) in the low temperature regime and (b) in the high temperature regime. To plot this figure, we use $\frac{\gamma}{\omega_0}=0.8$, $\frac{\gamma}{\omega_c}=1.6$, and $\frac{\omega_0}{\omega_c}=2.0$. }
\end{center}
\end{figure}  
Finally, specific heat of the system is given by 
\begin{eqnarray}
C_{QED}(T,B)&=&k_B\Big\lbrack\frac{4\pi^3(k_BT)^3(A_1+A_2+A_3)}{15\hbar^3\omega_0^6}\nonumber \\
&&+\frac{16\pi^5(k_BT)^5(B_1+B_2+B_3)}{21\hbar^5\omega_0^{10}}+....\Big\rbrack.
\end{eqnarray}
\subsection{QED : High temperature expansion ($k_BT>>\hbar\omega_0$)}
In this subsection, the high temperature thermodynamic propertis for the QED model are discussed in details. One can show that free energy for this model is 
\begin{eqnarray}
F_{QED}(T,B)=F_{Ohmic}^{HT}+\frac{\pi(k_BT)^2\gamma}{2\hbar\omega_0^2},
\end{eqnarray}
where $F_{Ohmic}^{HT}$ is the high temperature free energy for the Ohmic model (Eq. 33).
\begin{figure}[h]
\begin{center}
{\rotatebox{0}{\resizebox{12cm}{6cm}{\includegraphics{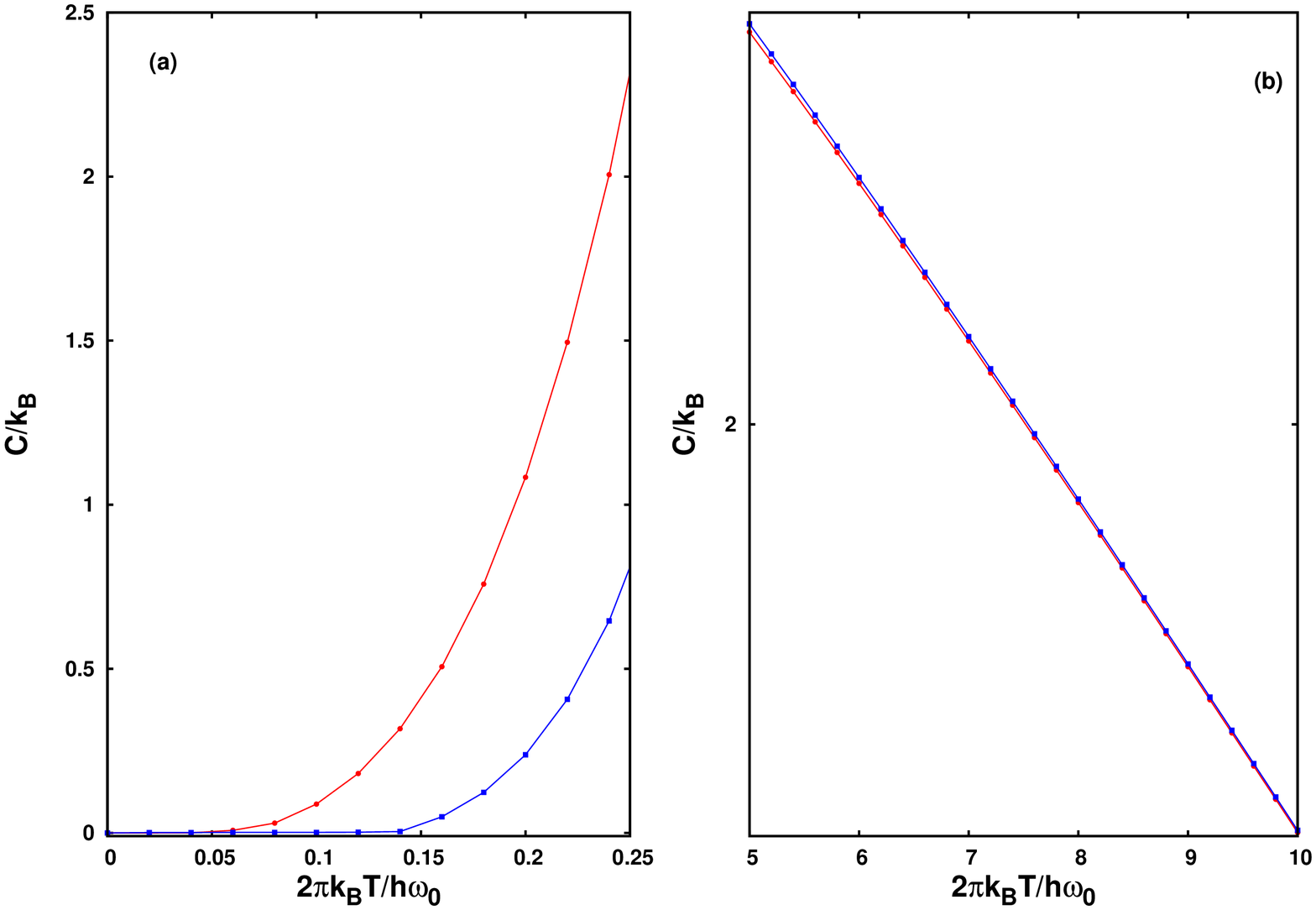}}}}
%\vskip-2.5cm
\caption{(color online)Plot of $\frac{C}{k_B}$ versus dimensionless temperature, $\frac{2\pi k_BT}{h\omega_0}$, for the charged magneto-oscillator (red filled circle) and for the free  charged oscillator (blue filled square) coupled to a blackbody radiation heat bath (a) in the low temperature regime and (b) in the high temperature regime. To plot this figure, we use $\frac{\gamma}{\omega_0}=0.8$, $\frac{\gamma}{\omega_c}=1.6$, and $\frac{\omega_0}{\omega_c}=2.0$. }
\end{center}
\end{figure} 
Entropy in the high temperature regime is given by
\begin{eqnarray}
\hskip-0.8cm
S_{QED}(T,B)=S_{Ohmic}^{HT}(T,B)-\frac{\pi(k_B^2T)\gamma}{\hbar\omega_0^2},
\end{eqnarray}
where, the high temperature specific heat for the Ohmic model ($S_{Ohmic}^{HT}(T,B)$) is given by equation (34). Internal energy of the system is given by 
\begin{eqnarray}
U_{QED}(T,B)=U_{Ohmic}^{HT}(T,B) -\frac{\pi(k_BT)^2\gamma}{2\hbar\omega_0^2},
\end{eqnarray}
where $U_{Ohmic}^{HT}(T,B)$ is given by equation (35).
Finally, specific heat of the system in the high temperature regime for the QED model is given by
\begin{eqnarray}
C_{QED}(T,B)=C_{Ohmic}^{HT}(T,B)-\frac{\pi k_B^2T\gamma}{\hbar\omega_0^2},
\end{eqnarray}
and $C_{Ohmic}^{HT}(T,B)$ is the specific heat for the Ohmic model in the high temperature regime (see equation 36).\\
\begin{figure}[t]
\begin{center}
{\rotatebox{0}{\resizebox{12cm}{8cm}{\includegraphics{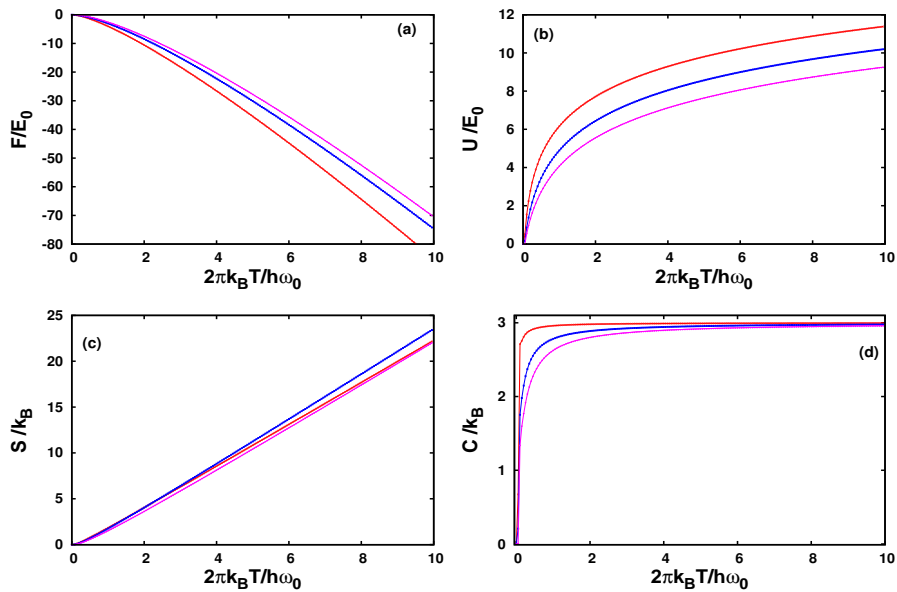}}}}
%\vskip-2.5cm
\caption{(color online) Plot of (a) $\frac{F}{E_{0}}$, (b)$\frac{U}{E_{0}}$, (c)$\frac{S}{k_{B}}$, and (d)$\frac{C}{k_{B}}$, versus dimensionless temperature, $\frac{2\pi k_BT}{h\omega_0}$, for the charged magneto-oscillator coupled to a Ohmic heat bath for the entire temperature regime with different values of dissipation parameter, $\gamma$, red filled circle ($\gamma/\omega_0=0.4$), blue filled square ($\gamma/\omega_0=0.8$), pink upward triangle ($\gamma/\omega_0=1.2$). To plot this figure, we also use $\frac{\omega_c}{\omega_0}=0.5$. }
\end{center}
\end{figure}
\begin{figure}[h]
\begin{center}
{\rotatebox{0}{\resizebox{12cm}{7cm}{\includegraphics{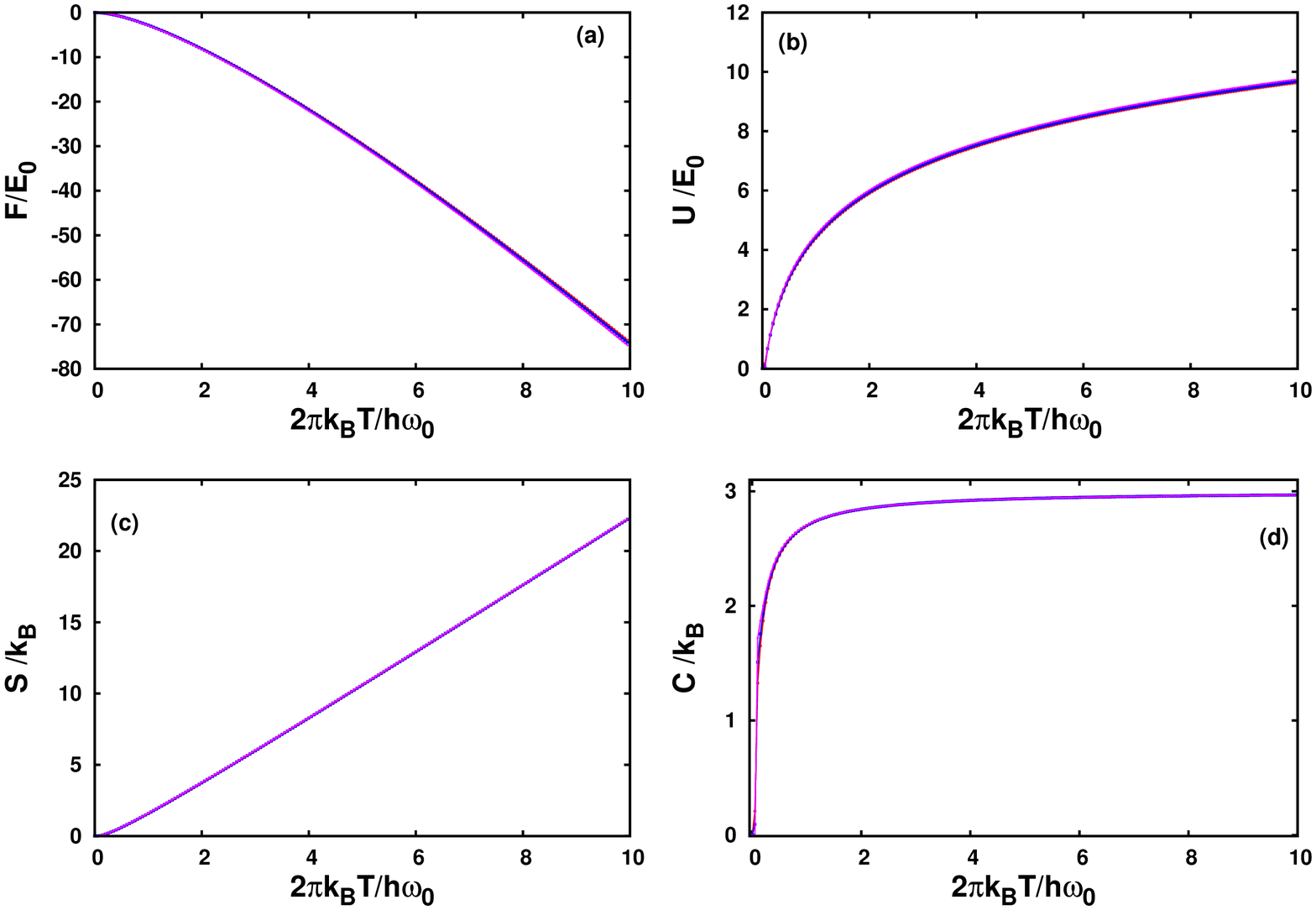}}}}
%\vskip-2.5cm
\caption{(color online) Plot of (a) $\frac{F}{E_{0}}$, (b)$\frac{U}{E_{0}}$, (c)$\frac{S}{k_{B}}$, and (d)$\frac{C}{k_{B}}$, versus dimensionless temperature, $\frac{2\pi k_BT}{h\omega_0}$, for the charged magneto-oscillator coupled to a Ohmic heat bath for the entire temperature regime with different values of cyclotron frequency, $\omega_c$, red filled circle ($\omega_c/\omega_0=0.25$), blue filled square ($\omega_c/\omega_0=0.5$), pink upward triangle ($\omega_c/\omega_0=0.75$). To plot this figure, we also use $\frac{\gamma}{\omega_0}=1.0$. }
\end{center}
\end{figure}
To explicitly demonstrate the distinguishing behaviour of different thermodynamic functions for three different heat bath models, we plot different thermodynamic quantities as a function of dimensionless temperature $(\frac{2\pi k_BT}{\hbar\omega_0})$ in different figures for the low temperature regime as well as for the high temperature regime. The results for different values of dissipation, $\gamma$, as well as for different values of $\omega_c$, are also analyzed here. We compare our results for various QTF of a charged magneto-oscillator with that of a free oscillator. From our analysis, one can conclude that the qualitative behaviour of different QTF for the charged magneto-oscillator in the high temperature regime is same as that of a free oscillator. Even, the quantitative values do not differ much from each other in the high temperature regime (see figures 1(b)-4(b) and figures 13(b)-16(b) ). In the low temperature regime, the free energy of the charged magneto-oscillator decays faster than that of the free oscillator for the Ohmic heat bath as well as for the QED model. On the other hand, the other QTF like internal energy ($U$), entropy ($S$), and specific heat ($C$) for the magneto-oscillator rises  faster than that of the free oscillator in the low temperature regime (see figures 1(a)-4(a) and figures 13(a)-16(a)). It has been observed that $\omega_c$ and $\gamma$ affect different QTF in the opposite manner. In the low temperature regime, $U$, $S$ and $C$ rise and $F$ decays much faster for higher values of $\gamma$ (see figure (5) and figure (6)) whereas they ($U$, $S$ and $C$) rise or $F$ decays much faster for lower values of $\omega_c$ (see figure (9) and figure (10)). In the high temperature regime, different QTF do not vary much for different values of $\omega_c$ (see figure (11) and figure (12)). But, the effect of different values of dissipation parameter, $\gamma$, is pronounced in the high temperature regime also. In this regime, $U$, $S$ and $C$ rise and $F$ decays much faster for higher values of $\gamma$ (see figure (7) and figure (8)). \\
We also plot general result for different quantum thermodynamic functions for the entire temperature regime by using equation (24) which is expressed in terms of Stieltjes J function (see figure (17) and figure(18)). For the numerical computation, we use the Lanczos formula for the Stieltjes $J$ function \cite{lanc1,press} :
\begin{equation}
\hskip-0.8cm
J(z)=(z+\frac{1}{2})\ln\frac{z+\gamma_1+1/2}{z}-\gamma_1-1/2+\ln\Big\lbrack d_0+\sum_{n=1}^N\frac{d_n}{z+n}\Big\rbrack, \ \ \ \ \Re{z}>0, 
\end{equation}
where $N=6$, $\gamma_1=5$, $d_0=1.00$, $d_1=76.18$, $d_2=-86.51$, $d_3=24.01$, $d_4=-1.23$, $d_5=0.001$, and $d_6=0.0$. It is seen that both at very high temperatures and at very low temperatures, the numerical results fairly match with the analytical expressions. Also, it is seen that the effect of dissipation parameter is pronounced in the entire temperature regime of our interest.\\
\section{Zero point energy}
For the sake of completeness, we are giving here zero point energy contribution of the charged magneto oscillator for the above mentioned three popular heat bath models.
It is known that free energy $F=U+TS$. Hence, the zero-point free energy is identical with zero-point energy. The zero-point free energy is obtained by replacing $f(\omega,T)\rightarrow \frac{\hbar\omega}{2}$ in Eq. (11) and in Eq. (12). Thus,
\begin{eqnarray}
F(0,B)&=&\frac{\hbar}{2\pi}\Big\lbrack 3\Big(\Omega\ln\Omega-\Omega^{\prime}\ln\Omega^{\prime}\Big)-\omega_1\ln\omega_1-\omega_1^*\ln\omega_1^*-\Omega_1\ln\Omega_1\nonumber \\
&&-\Omega_1^*\ln\Omega_1^*-\Omega_2\ln\Omega_2-\Omega_2^*\ln\Omega_2^*\Big\rbrack.
\end{eqnarray}
For single relaxation time model, the zero point free energy becomes
\begin{eqnarray}
\hskip-1.8cm
F(0,B)=\frac{\hbar}{2\pi}\Big\lbrack3\Big((\Omega^{\prime}+\gamma)\ln(\Omega^{\prime}+\gamma)-\Omega^{\prime}\ln\Omega^{\prime}\Big)-\gamma\ln\omega_0+2\omega_1^{\prime}\tan^{-1}\Big(\frac{2\omega_1^{\prime}}{\gamma}\Big)\nonumber \\
\hskip-1.8cm
-\lambda_1\ln(\lambda_1^2+\Gamma_1^2)-\lambda_2\ln(\lambda_2^2+\Gamma_2^2)+2\Gamma_1\tan^{-1}\Big(\frac{\Gamma_1}{\lambda_1}\Big)+2\Gamma_2\tan^{-1}\Big(\frac{\Gamma_2}{\lambda_2}\Big)\Big\rbrack,
\end{eqnarray}
where $\lambda_{1,2}=\frac{\gamma}{2}\pm\Big(\frac{b-a}{2}\Big)^{\frac{1}{2}}$, and $\Gamma_{1,2}=\frac{\omega_c}{2}\pm\Big(\frac{b+a}{2}\Big)^{\frac{1}{2}}$. On the other hand, the zero point contribution in the free energy for the Ohmic model is given by
\begin{eqnarray}
\hskip-1.8cm
F(0,B)&=&\frac{\hbar}{2\pi}\Big\lbrack3\gamma\Big(1-\ln(\omega_0\tau)\Big)+2\gamma\ln\omega_0-\lambda_1\ln(\lambda_1^2+\Gamma_1^2)-\lambda_2\ln(\lambda_2^2+\Gamma_2^2)\nonumber \\
\hskip-1.8cm
&&+2\omega_1^{\prime}\tan^{-1}\Big(\frac{2\omega_1^{\prime}}{\gamma}\Big)+2\Gamma_1\tan^{-1}\Big(\frac{\Gamma_1}{\lambda_1}\Big)+2\Gamma_2\tan^{-1}\Big(\frac{\Gamma_2}{\lambda_2}\Big)\Big\rbrack.
\end{eqnarray}
Finally, the zero point free energy for the QED model is given by
\begin{eqnarray}
F(0,B)&=&\frac{\hbar}{2\pi}\Big\lbrack 3\Big\lbrace\Big(\frac{\Omega^{\prime}\omega_0^2}{\omega_0^2+\gamma\Omega^{\prime}}\Big)\ln\Big(\frac{\Omega^{\prime}\omega_0^2}{\omega_0^2+\gamma\Omega^{\prime}}\Big)-\Omega^{\prime}\ln\Omega^{\prime}\Big\rbrace \nonumber \\
&&-\lambda_1\ln(\lambda_1^2+\Gamma_1^2)-\lambda_2\ln(\lambda_2^2+\Gamma_2^2)+2\omega_1^{\prime}\tan^{-1}\Big(\frac{2\omega_1^{\prime}}{\gamma}\Big)\nonumber \\
&&+2\Gamma_1\tan^{-1}\Big(\frac{\Gamma_1}{\lambda_1}\Big)+2\Gamma_2\tan^{-1}\Big(\frac{\Gamma_2}{\lambda_2}\Big)-\gamma\ln\omega_0\Big\rbrack.
\end{eqnarray}
\section{Conclusions}
Recent observations have suggested that small quantum systems are particularly sensitive to environmental effects \cite{mazenko,datta}. Dissipation and fluctuation effects often play a crucial role on the behaviour of such small systems. The influence of an external magnetic field on such small systems like quantum dots, quantum wires, and two dimensional electronic systems are of great interest in the fields of nanostructures. This fact has led to motivate us to study the quantum thermodynamic behaviour of a charged magneto-oscillator in an arbitrary heat bath at arbitrary temperature. Results for both high temperature and low temperature are shown explicitly in this paper.\\
\indent
We use the ``remarkable formula" of Ford et al \cite{ford4} for the free energy of the charged magneto-oscillator which involves single integral. This is an exact results for the free enrgy of a charged magneto-oscillator which takes into account interaction effects. Starting form this remarkable formula, we calculate several thermodynamic functions like free energy, entropy, internal energy, and specific heat for three different popular models like Ohmic, single relaxation time, and radiation heat bath or QED model. Explicit results are presented for the high temperature as well as for low temperatures. The central result of this analysis is that the qualitative behaviour as well as quantitative values of different thermodynamic quantities for the charged magneto oscillator differs from that of the free oscillator in the low temperature regime. The effect of $\omega_c$ and $\gamma$ are also investigated. They affect the rising or decay behaviour of different QTF just in the opposite manner to each other in the low temperature regime. In the high temperature regime, the effect of $\omega_c$ is not so significant. On the other hand, the effect of different $\gamma$ on the rising or decaying behaviour of different QTF is significant even in the high temperature regime. We have also shown the general results of various QTF by numerically evaluating equation (24). The numerical results obtained from the general expression fairly matches with the analytical expressions in the two limits of very high temperatures and very low temperatures. The experimentalist can make an estimation of change in values of different QTF from our study.\\
\indent
The application of this kind of analysis is manifold. Recent nanofabrication allows to create very small systems of few atoms in which environmental effects play an important role. This analysis is helpful in understanding environmental effects on small systems \cite{jacak}. The effect of magnetic field on the properties of two-dimensional quantum system in contact with a quantum heat bath is of great importance in the field of nanophysics \cite{jose}, quantum information theory \cite{bennett,zeilinger,giulini,myatt}, atomic and nuclear physics \cite{a,b}. In that sense, our study will be helpful in analyzing different thermodynamic properties  of  small quantum systems like quantum dots, quantum wires, two dimensional electronic systems in contact with a heat bath in the presence of a constant magnetic field. Recently Jordan and B$\ddot{u}$ttiker discussed about the entanglement energetics of quantum system at zero temperature \cite{buttiker}. So, following our method one can think of extending the study of entanglement enrgetics at finite temperature in the presence of an external magnetic field. Also Ratchov {\it et al} discussed about decrease in coherence length of Aharonov-Bohm like interferrometer due to the interaction with a zero temperature environment. Again, one can think of extending this work by following the method used in this paper for the finite temperature analysis \cite{ratchov}. Apart from that, one can think of another area in which thermodynamics plays a crucial role. Following the previous research of several authors \cite{bakenstein,bombelli,srednicki}, one can think of developing microscopic theory for the entropy of black holes. Thus, it is worthwhile to apply this approach to the study of thermodynamic properties of black holes. \\ 
{
\end{document}